\documentclass{vldb}
\usepackage{graphicx}
\usepackage{epstopdf}
\pdfoutput=1

\usepackage{amsthm}
\usepackage{balance}  
\usepackage[ruled,linesnumbered,noend,vlined]{algorithm2e}
\usepackage{color,soul}
\usepackage{subfig}
\usepackage{enumitem}
\usepackage{tabu}
\setitemize{leftmargin=10pt, noitemsep,topsep=0pt,parsep=0pt,partopsep=0pt}
\DeclareMathSizes{10}{7}{8}{8}

\vldbTitle{}
\vldbAuthors{}
\vldbDOI{}
\vldbVolume{}
\vldbNumber{}
\vldbYear{}

\begin{document}


\title{Finding Effective Geo-Social Group for Impromptu Activity with Multiple Demands}



%
%
%
%
\numberofauthors{1}
\author{
    \alignauthor
Lu Chen\textsuperscript{$\dagger$}, Chengfei Liu\textsuperscript{$\dagger$}, Rui Zhou\textsuperscript{$\dagger$}, Jiajie Xu\textsuperscript{$\S$}, Jianxin Li\textsuperscript{$\P$} \\
\textsuperscript{$\dagger$}\affaddr{Swinburne University of Technology},
\textsuperscript{$\P$}\affaddr{Soochow University},  \textsuperscript{$\S$}\affaddr{Deakin University}\\
\email{\textsuperscript{$\dagger$}\{luchen, cliu, rzhou\}@swin.edu.au \textsuperscript{$\S$}xujj@suda.edu.cn \textsuperscript{$\P$}jianxin.li@deakin.edu.au
}}

\maketitle

\begin{abstract}
Geo-social group search aims to find a group of people proximate to a location while socially related.
One of the driven applications for geo-social group search is organizing an impromptu activity.
This is because the social cohesiveness of a found geo-social group ensures a good communication atmosphere and the spatial closeness of the geo-social group reduces the preparation time for the activity.
Most existing works treat geo-social group search as a problem that finds a group satisfying a single social constraint while optimizing the spatial proximity.
However, when an impromptu activity has additional demands on attendees, e.g., the activity requires that the attendees have certain set of skills, the existing works cannot find an effective geo-social group efficiently.
In this paper, we study how to find a group that is most proximate to a query location while satisfying multiple constraints.
Specifically, the multiple constraints on which we focus include social constraint, size constraint and keyword constraint.
We propose a novel search framework which first effectively narrows down the search space with theoretical guarantees and then efficiently finds the optimum result.
Although our model considers multiple constraints, novel techniques devised in this paper ensure that search cost is equivalent to parameterized constant times of one time social constraint checking on a vastly restricted search space.
We conduct extensive experiments on both real and semi-synthetic datasets for demonstrating  the efficiency of the proposed search algorithm.
To evaluate the effectiveness, we conduct two case studies on real datasets, demonstrating the superiority of our proposed model.

\end{abstract}

\everymath{\small}

\newtheoremstyle{exampstyle}
{3pt} 
{3pt} 
{\itshape} 
{} 
{\bfseries}
{.} 
{.5em} 
{} 
\theoremstyle{exampstyle}

\newtheorem{theorem}{Theorem}
\newtheorem{lemma}{Lemma}

\newtheorem{example}{Example}
\newtheorem{property}{Property}
\newcommand{\qk}{\varphi}
\newtheorem{definition}{Definition}

\newcommand{\spatialG}{\textsf{MKASG}}
\newcommand{\spatialGspace}{\textsf{MKASG~}}

\newcommand{\algbaselineA}{\textsf{MKASG-}}
\newcommand{\algbaselineB}{\textsf{MKASGDec}}
\newcommand{\algbaselineC}{\textsf{MKASGBinInd}}
\newcommand{\algbaselineD}{\textsf{MKASGInc}}

\newtheorem*{reproblem}{Research Problem}

\SetAlFnt{\small\normalfont}
\SetAlCapHSkip{0em}
\SetAlgoSkip{}

\setlength\floatsep{1.25\baselineskip plus 3pt minus 3pt}

\setlength\textfloatsep{1.25\baselineskip plus 3pt minus 3pt}
\setlength\intextsep{1.25\baselineskip plus 3pt minus 3pt}

\setlength{\abovecaptionskip}{2pt}
\setlength{\belowcaptionskip}{-12pt}

\section{Introduction}
As the geo-social networks become popular, finding geo-social groups has drawn great attention in recent years.
In general, geo-social group search problem~\cite{liu2012circle, armenatzoglou2013general,zhu2017geo, 7202872} aims to find a group that is socially cohesive while spatially closest to a location, i.e., the found group satisfies a single social constraint while optimizing a distance objective function for most works.
This is different from most social-aware spatial search works~\cite{wu2014social, li2014spatial, ahuja2015geo,  Ghosh:2018:FSS:3282495.3302535} that consider various objectives together as an aggregate objective function and find the result that is optimum w.r.t. the aggregate function.
One of the most motivating applications for geo-social group search is instant organization of impromptu activities.
This is because two nice proprieties of geo-social groups. Firstly, the social cohesiveness of a geo-social group ensures the members are socially close within the group, which is key to ensure a good communication atmosphere for the activity.
Secondly, subjecting to social cohesiveness, a geo-social group is the one that is closest to the location of the activity, which reduces the waiting time for the activity potentially.
However, since most of the existing geo-social group studies only focus on social constraint while optimizing the spatial closeness, they become less useful when an activity has more demands, e.g., demanding attendees with certain skills and demanding minimum number of attendees.
Let us consider one of the application scenarios below.

\begin{figure}[t]
	\centering
	\subfloat[\noindent Graph data]{\includegraphics[height=40mm]{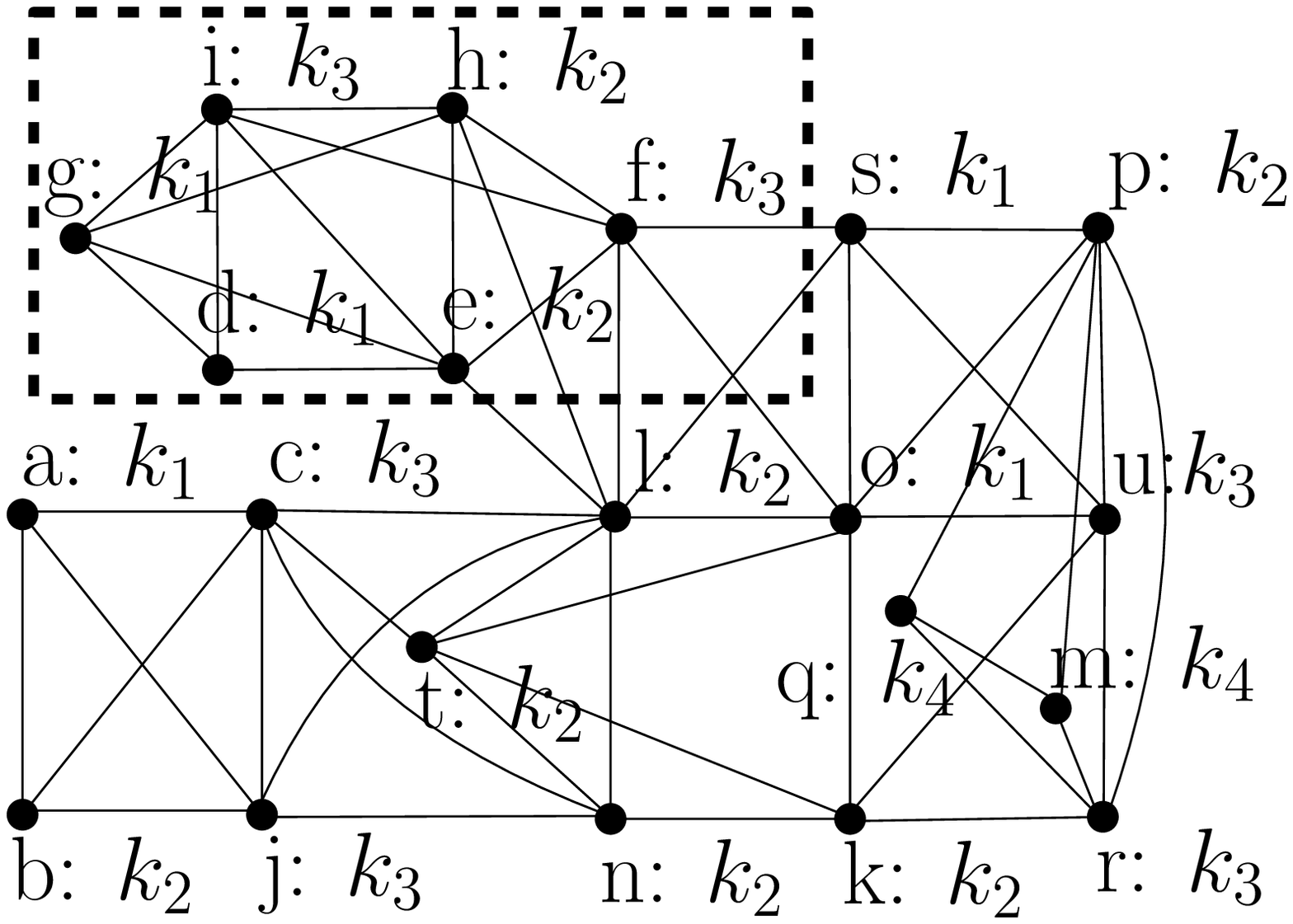}}
	\hspace{-0.2cm}
	\subfloat[\noindent Spatial distribution]{\includegraphics[height=40mm]{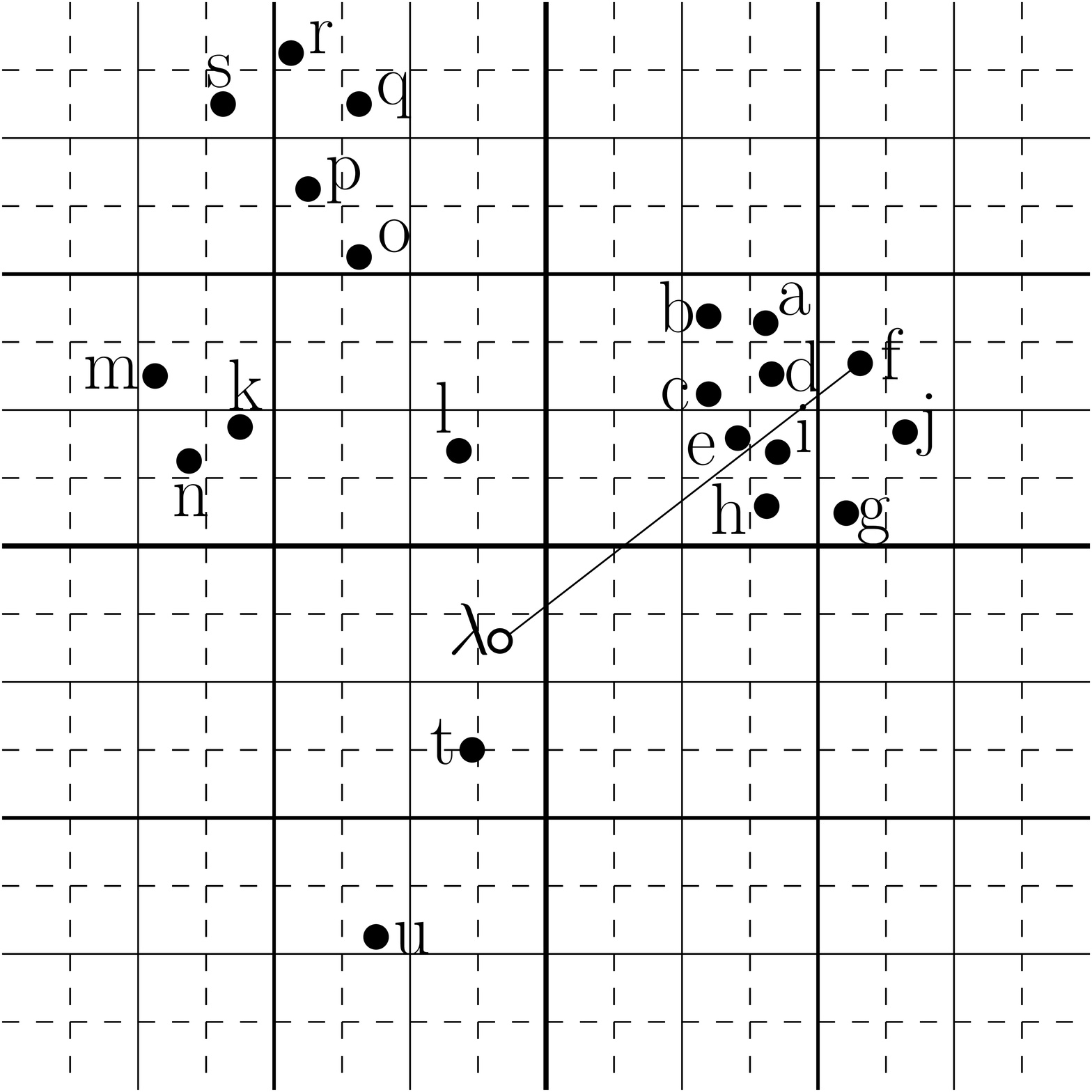}}	
	\caption{Graph with location and keyword}\label{fig:sag}
	\vspace{-5pt}
\end{figure}

\textit{Online open-world game data: finding participants for a real time quest}. For online open-world game data, each player is associated with a friend list, an attribute describing the role of a player, and location information showing his/her location in the virtual world.
Suppose there is a real time quest requested in a randomly location with duration of $15$ minutes.
The quest has a set of suggested roles and suggests that each role shall have no less than $2$ players for accomplishing the quest.
The gaming system would like to formulate a group of participants who are adequate to carry out the quest.
Who shall be the players in the group?

To effectively find the desired geo-social group for the above scenario, extra factors shall be considered thoroughly in addition to social and spatial closeness, i.e., the minimum number of players for each suggested role.
If there are more demands, the effort to coordinate them increases substantially.
As such, it is imperative to devise efficient novel techniques to alleviate the effort for planning or organising activities with multiple demands.
A specific motivating example is shown below.


\begin{example}\label{ex:mv}
Figure~\ref{fig:sag} illustrates the gaming data, which consists of graph data in Figure~\ref{fig:sag}(a) and spatial data in Figure~\ref{fig:sag}(b), when a real time quest is happening at the location labelled as $\lambda$.
The graph data contain friendships for players and the current role of players in terms of keyword. The spatial data contain the current location for each users.
Let the quest has a suggested role of $\{k_{1}, k_{2}, k_{3}\}$ and has a suggested minimum number of players, say, $2$ for each suggested role.
Below, we show the desired result and the results found by the most related models.
\end{example}

The desired group for accomplishing the quest is the subgraph enclosed by the dashed rectangle in Figure~\ref{fig:sag}(a).
The players found within the group have strong relationship while preserving spatial proximity to the query location $\lambda$.
Simultaneously, the group contains players satisfying all roles recommended by the quest and the group also has sufficient players, i.e., two players for each suggested role. Considering a single social constraint (e.g., $k$-core~\cite{KhaouidKCore}), geo-social group works such as~\cite{zhu2017geo} tend to find the nearest group satisfying the social constraint, i.e., $\{a,c,b,j\}$ induced subgraph in Figure~\ref{fig:sag}(a).
Considering a social constraint and the exact group size constraint, existing works such as~\cite{7202872, Ghosh:2018:FSS:3282495.3302535} are likely to find $\{a,b,c,j,l,t\}$ induced subgraph in Figure~\ref{fig:sag}(a).
None of them can find the group as the desired one since they do not consider that the activity has multiple demands as discussed above.

\noindent\textbf{Geo-social group with multiple constraints}. The example motivates us to study a novel type of geo-social group search problems for impromptu activities with multiple demands, and propose efficient solutions.
In particular, the model which we study finds a group with multiple constraints induced by various demands of an activity while preserving that the group is most spatially proximate to the activity location, in which the spatial proximity is measured by the distance of the person in the group that is most distant to an activity location.
The multi-constraint geo-social group search problem that we focus on is to find \spatialGspace - a \textsf{G}roup of people with \textsf{M}inimum requirements of \textsf{K}eyword cohesiveness, \textsf{A}cquaintance (social strength) and size while preserving its \textsf{S}patial-proximity to a given location.
We name this problem as \spatialG~search problem.

\noindent\textbf{Existing search framework}. 
Most existing approaches~\cite{liu2012circle, armenatzoglou2013general,zhu2017geo} for finding a geo-social group mainly based on the nearest neighbour search framework.
This framework progressively adds vertices that potentially satisfy social constraint according to nearest neighbour order (w.r.t. the activity location), while checking the social constraint after each vertex is added. It returns the optimum result when it finds a subgraph satisfying the social constraint for the first time.
This framework is efficient when considering a single social constraint.
When coming to geo-social group with multiple constraints, this framework becomes less attractive since some constraints, e.g., minimum size constraint discussed above, may enlarge the size of desired geo-social group.
This makes the times of multi-constraint checking substantially large, resulting poor performance.


\noindent\textbf{Challenges}. As discussed above, a general effective framework for searching geo-social group with multiple polynomial checkable constraints is required in urgency.
This arises challenges as follows.
Firstly, can we have a search framework that can narrow the search space fast while preserving the correct result?
Secondly, can we have a theoretical bound for the size of the narrowed search space?
Thirdly, given the specific constraints in \spatialG, can we reduce the time complexity of multi-constraint checking approach to constant times of single constraint checking?

\noindent\textbf{Our approach}. In this paper, we devise a novel search framework for effectively finding geo-social group with multiple constraints.
This search framework contains expanding and reducing stage.
The expanding stage addresses the first two challenges. It approaches to a search space that is sufficient large to contain the optimum result at a cost equivalent to \textit{constant times} of the time complexity of multi-constraint checking.
The approached search space is no greater than the size of the optimum search space with a ratio of parameterized constant, which vastly restricts the search space for the reducing stage.
For the reducing stage, we adapt the method proposed in~\cite{liinfluentalcom2015}.
For \spatialG~search problem, within the proposed search framework, we further devise novel techniques including keyword aware truss union and keyword aware spanning forest, which reduce the overall search complexity, including expanding and reducing stages, to constant times of the social constraint checking.
This addresses the third challenge.
We also propose novel pruning techniques that further improve the search performance as much as possible.

\noindent\textbf{Contribution}. Our predominant contributions in this paper are summarised as follows.

\begin{itemize}
  \item We study finding geo-social group with multiple constraints, considering minimum keyword, social acquaintance and size constraints while preserving its spatial proximity to a specific query location. \hfill(\textbf{Section}~\ref{sec:pf})
  \item We devise an effective search framework for multiple constraints geo-social group search problem, which first approaches to the region containing a group stratifying all constraints and then reduces the group to \spatialG~to guarantee the spatial proximity.  \hfill(\textbf{Section}~\ref{sec:framework})
  \item For the expanding stage, we propose a power law based expanding strategy which ensures that the evaluated search space of the expanding range is restricted. We further propose effective techniques including search region lower bound, and keyword aware truss union-find operation to speed up this stage.  \hfill(\textbf{Section}~\ref{sec:epanding})
  \item For the reducing stage, we propose novel keyword frequency aware spanning forest, which guarantees the total cost of the reducing stage to its lower bound for \textsf{MKASG} search.  \hfill(\textbf{Section}~\ref{sec:reducing})
  \item We conduct extensive experiments on real datasets to demonstrate the efficiency and effectiveness of the proposed algorithm and geo-social group model. \hfill(\textbf{Section}~\ref{sec:exp})
\end{itemize}

\section{Problem formulation}\label{sec:pf}

In this section we formulate \spatialG~with social, keyword and size constraints and \spatialG~search problem.
Some of other constraints on geo-social group that can be solved by our proposed method will be discussed in Section~\ref{sec:warp}.

\noindent\textbf{Data}. We model data with network structure, spatial attribute and textual attribute as an undirected graph $G=(V,E)$.
$G$ has a set of vertices (users) $V$ and a set of edges (friendships) $E$.
For each vertex $v\in V(G)$, $v$ has a piece of location information expressed as latitude and longitude denoted as $(v.x,v.y)$, and has a keyword denoted as $v.A$ that describes the current role of $v$.

We formally define the query for searching \spatialG.

\noindent\textbf{Query for \spatialG}. We allow users to give a query $Q$ consisting of a query location $\lambda$,
a set of keywords $\qk$ that describe the roles of the desired group members,
an integer parameter $\rho$ for defining minimum size of the group,
and an integer parameter $c$ that defines social cohesiveness.

\noindent\textbf{Multiple constraints for \spatialG}. Now we define the multiple constraints of \spatialG, given an \textsf{MKASG} query.

\noindent\textit{\underline{Social constraint}}. We consider minimum trussness to measure the social cohesiveness of an \spatialGspace $S\subseteq G$.
Trussness is defined based on the number of triangles each edge is involved in a graph.
In general, given a subgraph $S\subseteq G$, we use $\bigtriangleup_{uvw}$ to denote a triangle consisting of vertices $u,v,w\in V(S)$.

\textit{Support}. The support of an edge $e(u,v)\in E(S)$, denoted by $sup(e,S)$, is the number of triangles containing $e$, i.e., $sup(e,S)=|\{\bigtriangleup_{uvw}:w\in N(v,S)\cap N(u,S)\}|$, where $N(v,S)$ and $N(u,S)$ are the neighbours of $u,v$ in $S$ correspondingly.

\textit{Minimum subgraph trussness}. The trussness for a subgraph $S$ is defined as an integer $c$ that is $2$ plus the minimum possible support for edges in $E(S)$. That is, the minimum subgraph trussness defines that for every edge $e\in E(S)$, the number of triangles in which $e$ participates shall be no less than $c$ - $2$.

Based on the definition of trussness, we define the $c$-truss constraint of an $\spatialG$~$S$ as follows:
\begin{definition}\textup{\textbf{$c$-truss constraint}}.
An \spatialG~$S$ satisfies $c$-truss constraint if the trussness of $S$ is $c$, and $S$ is connected.
\end{definition}

Intuitively, if $S$ satisfies $c$-truss constraint, the vertices of an edge in $S$ have at least $c$-$2$ common neighbours in the group $S$, every vertex in $S$ has no less than $c$-$1$ edges and at least $c$-$1$ edges have to be deleted in order to make $S$ disconnected. An $S$ with a large value $c$ indicates strong internal social relationships over vertices.

\textit{Example}. For instance, in Figure~\ref{fig:sag}(a), the whole graph is a $4$-truss. Every edge in this graph involves no less than $2$ triangles. 

\noindent\textit{\underline{Keyword constraint}}.
We adopt the concept of collective keyword coverage to measure the keyword cohesiveness between the keyword attributes of $V(S)$ and query keywords $\qk$.

\textit{Collective keyword coverage}. Given a group $S$ and the query keywords $\qk$, the attributes of $V(S)$ collectively cover $\varphi$ if and only if $\cup_{v\in V(S)} v.A = \varphi$.

\noindent\textit{\underline{Minimum size constraint}}.
In real application, we could allow users to specify the minimum size of the group directly. 
However, this is likely to result in that the attributes of the found group members overemphasize on part of $\qk$, which is undesired.
To mitigate such effect, we propose an alternative approach defining the minimum size of the group together with the keyword constraint.
We introduce the definition of minimum $\rho$ keyword vertex constraint. 

Given a set of keyword $\qk$ $=$ $\{k_{1},\ldots, k_{|\qk|} \}$, a social group $S$, and let $V(S_{k_{i}})\subseteq V(S)$ be the set of vertices in $V(S)$ containing keywords $k_{i}\in \qk$, the minimum $\rho$ keyword vertex constraint is defined as follows.

\begin{definition}\textup{\textbf{Minimum $\rho$ keyword vertex constraint}}
	Given an integer $\rho$, $\qk$ and $S$, $S$ satisfies minimum $\rho$ keyword vertex constraint if: $  min\{|S_{k_{i}}||\forall k_{i}\in \qk, S_{k_{i}} \subseteq S\} \ge \rho$.

\end{definition}

With the minimum $\rho$ keyword vertex constraint, the size of a group is no less than $\rho \times |\qk|$.
In the following of this paper, we call minimum $\rho$ keyword vertex constraint as keyword vertex constraint.

\noindent\textbf{Searching objective for \spatialG~search}. Now, we formalize the spatial proximity measurement for \spatialGspace and the research problem studied in this paper.

\noindent\textit{\underline{Spatial proximity}}.
Given a query location $\lambda$, we consider a distance function to measure the closeness between $\lambda$ and an \spatialG~$S$ as:

\begin{definition}\label{def:spatialPro} \textup{Distance measurement}.
$$
dist(\lambda,S) = max\{\lVert\lambda-v\rVert|v\in V(S)\},
$$	
\end{definition}
\noindent where $\rVert \lambda-v \lVert$ denotes Euclidean distance between $v$ and $\lambda$.


\begin{definition}\label{def:rhottruss} \textup{\textbf{$(\rho,c, d)$-truss}}.
	Given a $Q=\{\lambda,\rho,\varphi, c\}$ and a distance threshold $d$, a subgraph $S\subseteq G$ is a $(\rho,c,d)$-truss, if it satisfies all the constraints below.
	\begin{itemize}
		\item $min\{|V(S_{k_{i}})||\forall k_{i}\in \qk, V(S_{k_{i}}) \subseteq V(S)\}\ge \rho$.
		\item $S$ satisfies $c$-truss constraint.
        \item $dist(\lambda,S)$ $\le$ $d$.
	\end{itemize}
\end{definition}

\begin{reproblem}  \textup{\textbf{\spatialG~search}}.
	Given a query $Q=\{\lambda,\rho,\varphi, c\}$ and $G$, return $(\rho,c, d)$-truss $S^{*}$ so that there is no $(\rho,c, d')$-truss $S'$ with $d'\le d$.

\end{reproblem}

\noindent\textbf{Example}. Come back to Example~\ref{ex:mv}, and set a query for \spatialG~search with $\lambda$, $\qk=\{k_{1},k_{2},k_{3}\}$, $\rho =2$, $c=4$. \spatialG, denoted as $S^{*}$, is the subgraph in the doted area.
It is a $4$-truss subgraph, and for every keyword in $\qk$ there are no less than two members whose attributes match the keyword.
It is also the group closest to $\lambda$ subject to the social, keyword and size constraints.

\vspace{-10pt}
\section{Baseline Solutions}

In this section, we discuss three baseline solutions that find the exact result.

\noindent\textbf{Incremental approach}. Given a query, this approach progressively includes a vertex into a candidate set according to nearest neighbour order w.r.t. the query location.
Every time a vertex is added into the candidate set, this approach checks if there is a subgraph induced by vertices in the candidate set that satisfies all constraints.
If there is one, the approach stops and returns the subgraph as result. Otherwise, this approach keeps on exploring the vertices in order.

This method has a time complexity of $\mathcal{O}$ $($ $|V(G)|$ $|E(G)|^{1.5}$$)$.
The dominated cost is induced by repeatedly checking $c$-truss constraint and keyword vertex constraint. 

\noindent\textbf{Decremental approach}. Borrowing the technique proposed in~\cite{liinfluentalcom2015}, a baseline with better time complexity can be derived.
This approach progressively deletes the vertex most distant to $\lambda$. When a most distant vertex is deleted, this approach further deletes edges that do not satisfy trussness constraint. This ensures that every time before deleting the next most distant vertex, the remaining subgraphs are still $c$-truss.
To adapt this approach for our problem, after trussness checking, for the remaining truss subgraphs we further check if there is connected $c$-truss satisfying both size and keyword vertex constraint using depth-first search. The decremental approach progressively deletes the most distant vertex and performs the multi-constraint checking until there is no subgraph that satisfies all constraints simultaneously.
The last subgraph that satisfies all constraints becomes the result.

The time complexity of this approach is $\mathcal{O}$ $(|V(G)||E(G)|$ $+$ $|E(G)|^{1.5})$. This approach can reduce the cost of truss checking. But, it suffers from exploring large search space.

\noindent\textbf{Binary search based approach}.
This approach progressively guesses a distance $d$ via binary search. For a distance $d$, this approach checks if there is a subgraph that satisfies all constraints in the subgraph induced by vertices having distance no greater than $d$ to the query location $\lambda$.
If there is one, this approach reduces the $d$ to $\frac{d}{2}$ and continues. If there is no such a subgraph, this approach increases $d$ to $\frac{d'-d}{2}$ where $d'$ is the last evaluated distance and checks the corresponding subgraphs.
For any two consecutive evaluated $d'$ and $d$, if there is no vertices having distance to $\lambda$ between $d'$ and $d$, the search stops and the last subgraph satisfying all constraints becomes the result.
To support retrieve subgraphs based on $d$ efficiently, we use R-tree index in this method.


The time complexity of this approach is $\mathcal{O}$ $(\log_{2}(|V(G)|)$ $|E(G)|$ $+$ $\log_{2}(|V(G)|)|E(G)|^{1.5})$. The major drawback of this approach is that its search space is large even though it can approach to the optimum result fast.

\noindent\textbf{Discussion}. The advantage of incremental approach is if the result is near to the query location, the search space is quite restricted. The advantage of the decremantal approach is that it can reduce the cost of truss computation. The advantage of binary search based approach is it can quickly approach to optimum result in the worst case. Clearly, an ideal search framework shall take all the advantages. This motivate us to devise a novel framework that only explores restricted area, approaches to the optimum result fast and reduces multi-constraint checking as much as possible.


\begin{figure}[t]
\vspace{-10pt}
	\centering
	\subfloat[\noindent Spatial distribution]{\includegraphics[height=36mm]{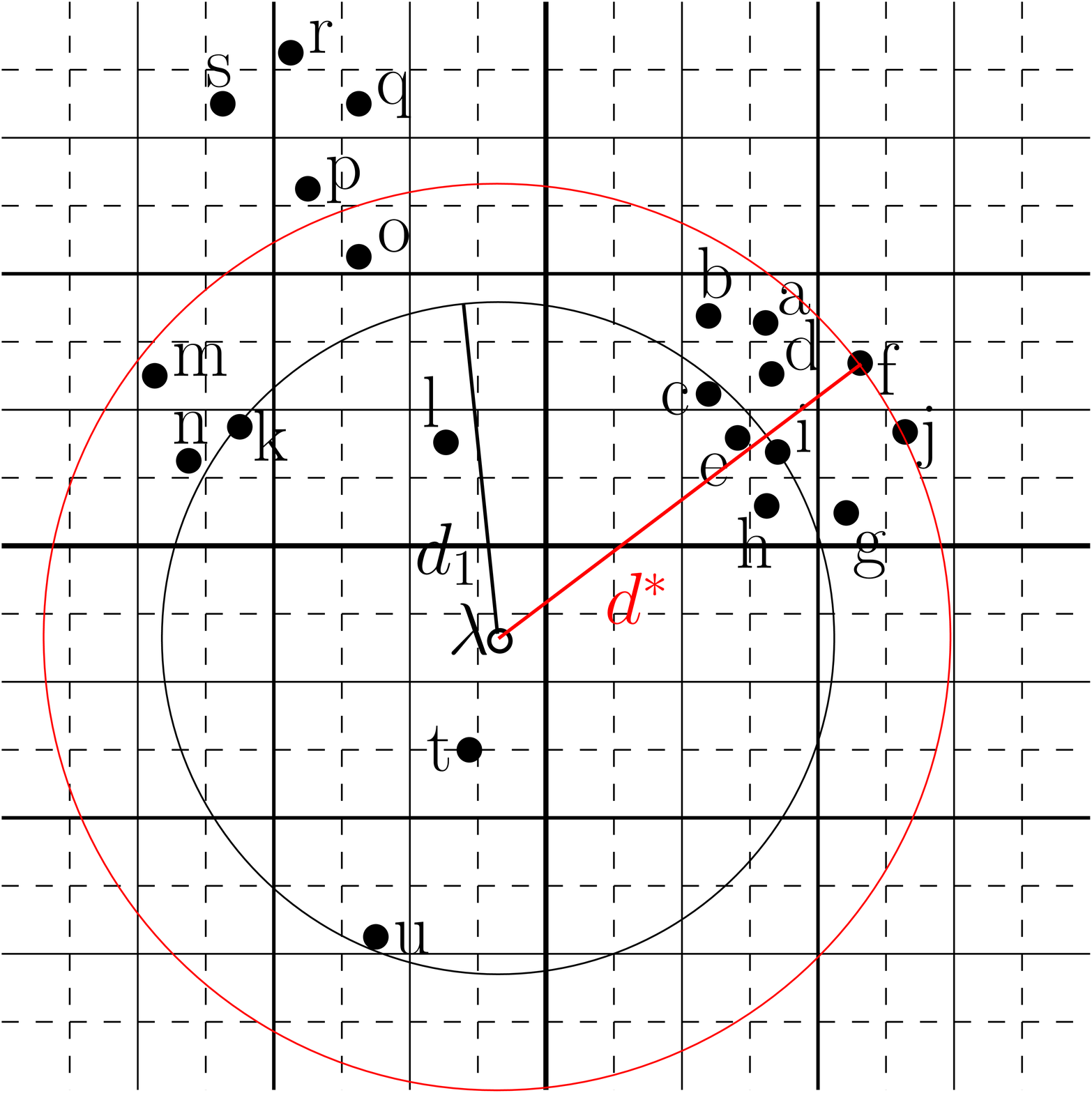}}
	\hspace{-0.2cm}
	\subfloat[\noindent $H_{\le d_1}$ and $H_{\le d^{*}}$]{\includegraphics[height=40mm]{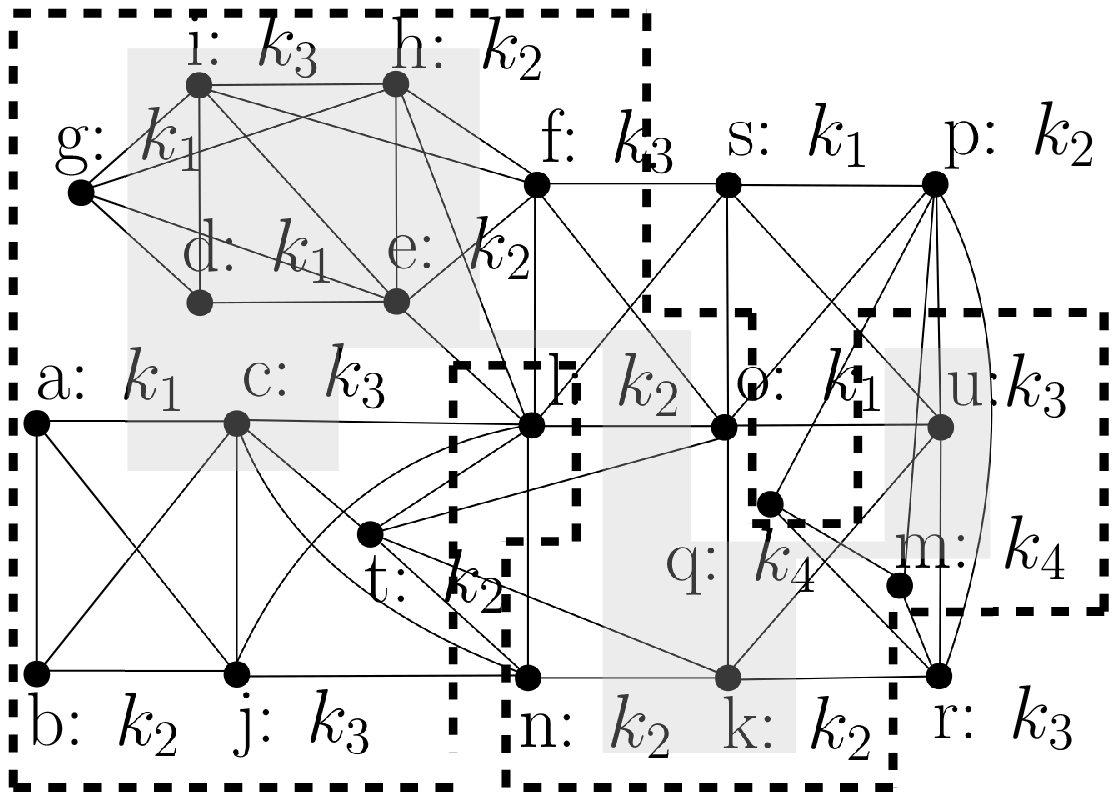}}	
	\caption{Graphs for $H_{\le d_{1}}, H_{\le d^{*}} $}\label{fig:ledgraph}
\end{figure}

\vspace{-10pt}
\section{Search Framework}\label{sec:framework}


Before showing the search framework, we firstly introduce a pre-pruning technique and some definitions.


\noindent\textbf{Maximal ($\rho$, $c$)-truss based pruning}. A maximal ($\rho$, $c$)-truss is a ($\rho$, $c$, $d$)-truss that cannot be extended by adding either an edge or a vertex while considering $d$ as $\infty$.

Given an \spatialG~query containing parameters $\rho$ and $c$, it is clear that \spatialG~for the query can only reside in a maximal ($\rho$, $c$)-truss if it exists. As such, given the \spatialG~query and $G$, computing maximal ($\rho$, $c$)-truss subgraphs contained in $G$ would reduce the search space significantly. This can be done by traversing maximal $c$-truss subgraph with the state of the art truss technique~\cite{DBLP:conf/sigmod/0001Y19}.

\begin{definition}\textup{\textbf{$d$ radius bounded graph}}.
Given a query location $\lambda$, a subgraph $H$ and a distance threshold $d$, $d$ radius bounded graph, denoted as $H_{\le d}$, is the subgraph of $H$ induced by vertices of $H$ with distance to $\lambda$ no greater than $d$.
\end{definition}

We would like to highlight a special instance of $d$ radius bounded graph, $d^{*}$ radius bounded graph ($H_{\le d^{*}}$), which is the $d$ radius bounded graph just large enough to contain \spatialG~for a query, i.e., there is no $H_{\le d'}$ such that $H_{\le d'}$ contains \spatialG~and $d'$ $<$ $d^{*}$. We refer $H_{\le d^{*}}$ as optimum search space since it is just large enough to contain \spatialG~for the query.

For instance, in Figure~\ref{fig:ledgraph}, $H_{\le d_{1}}$ and $H_{\le d^{*}}$ are demonstrated. $d_{1}$ and $d^{*}$ identified regions are displayed in Figure~\ref{fig:ledgraph}(a), i.e., cycles centred by $\lambda$ with radius of $d_{1}$ and $d^{*}$ respectively.
The subgraphs are shown in Figure~\ref{fig:ledgraph}(b), i.e., $H_{\le d_{1}}$ is the subgraph in doted area and $H_{\le d^{*}}$ is the subgraph in grey coloured area.
$H_{\le d^{*}}$ is the optimum search space containing \spatialGspace for the query in Example~\ref{ex:mv}.

Next we show the search framework for \spatialG. It firstly approaches to a $H_{\le d'}$ just sufficient large to constrain $H_{\le d^{*}}$ quickly. Then it reduces $H_{\le d'}$ to the optimum result.



\begin{algorithm}[t]
\scriptsize
	\KwIn{$H$, $Q$}
    \KwOut{$S^{*}$}
	$d \leftarrow$ initial search distance for $H_{\le d}$ \;
	$d^{*}\leftarrow \infty$, $S^{*}\leftarrow$ $\emptyset$ \;

    \tcc{Expanding stage}
	$S \leftarrow \textsc{ispcTrussIn}$($H_{\leq d}$) \;
	\While{ $S$ is  $\emptyset$}{
		$H_{\le d'}$ $\leftarrow$ \textsc{newRange}($d$)\;

		$S \leftarrow$ \textsc{ispcTrussIn}($H_{\leq d'}$ )\;
		$S^{*}\leftarrow S$, $d\leftarrow d^{'}$\;
	}

    \tcc{Reducing stage}
	$S^{*} \leftarrow$ \textsc{redcuepcTruss}($S^{*}$)\;
	\Return $S^{*}$\;
	
	\caption{\textsc{searchMKASG}($Q$,$H$)}\label{alg:eralg}
\end{algorithm}

\noindent\textbf{The framework}. As shown in Algorithm~\ref{alg:eralg}, \spatialGspace search framework consists of two stages: expanding stage (lines 3 to 7) and reducing stage (line 8). During the expanding stage,  Algorithm~\ref{alg:eralg} intends to quickly identify $H_{\le d}$ that is just sufficiently large to contain the optimum search space $H_{\le d^{*}}$ by exploring  $H_{\le d}$ that progressively gets larger, in which \textsc{isptTrussIn} is called to determine the existence of a subgraph satisfying all constraints.
For the reducing stage, to get the optimum result, \textsc{reducepcTruss} attempts to progressively remove the vertex that is the most distant to $\lambda$ in $S^{*}$.
The last survived ($\rho$, $c$)-truss during the vertices removing process is the optimum result.

In the following sections, we will discuss details of the two stages. We will propose techniques that make expanding stage having the time complexity of one time calling of \textsc{isptTrussIn}.
For the reducing stage, we will propose novel online index and combine the index with our proposed reducing strategy to efficiently check all constraints of \spatialG.
Eventually, our proposed techniques can guarantee that Algorithm~\ref{alg:eralg} has a time complexity of one time truss computation.

\section{Expanding Stage}\label{sec:epanding}
In this stage,  we explore a set of $d$ radius subgraphs, starting from a relatively small $d$ radius subgraph and stopping at the first $d$ radius subgraph that is a super graph of $H_{\le d^{*}}$.


\noindent\textbf{Challenges}. Since expanding stage involves expensive constraint checking, our first challenge is how to devise an expanding strategy that can elegantly bound the overall computations tightly?
On the other hand, if we can expand to $d^{*}$ with less number of attempts, the performance will be improved.
This can be achieved by starting the search from a $d$ radius graph with $d$ that is close to but no greater than $d^{*}$. This arises the second challenge: can we identify such initial search range efficiently?
At last, when processing an $H_{\le d}$ during the expanding stage, if we apply multi-constraint checking just on some restricted subgraphs of $H_{\le d}$ that potentially contain a ($\rho$, $c$)-truss, the search performance can be further boosted. This arises the third challenge on how to quickly identify those potential subgraphs in $H_{\le d}$?

In the following sub-sections, we will address these three challenges consecutively.

\subsection{Expanding Strategy}
In this part, we propose an expanding strategy which can bound the total amount of subgraphs that will be evaluated.

We first define an expanding invariant as follows. 

\begin{definition}\textbf{$\Delta$ size invariant}.
    Let $\{d_{1}, d_{2}, \ldots, d_{i}\}$ be the series of radius for defining $d$ radius graphs, for any two consecutive $d$, $d'$ in the series, we define $\Delta$ invariant as
    $$
        \Delta = \frac{|E(H_{\leq d'})|}{|E(H_{\leq d})|},
    $$
    in which $\Delta > 1$ must hold.
\end{definition}


\noindent\textbf{The strategy}. The strategy applied for the expanding stage is to maintain $\Delta$ size invariant over any two consecutively evaluated $H_{\le d}$, $H_{\le d'}$. Applying $\Delta$ invariant for expanding stage guarantees two nice properties below.

\begin{property}\textbf{Nearest first search}.
 Vertices accessed by the expanding stage are in non-increasing order according to their distance to $\lambda$ on a batch basis.
\end{property}

\begin{property}\textbf{Power law expansion}~\cite{bi2018optimal}.
The sizes of the set of $d$ radius graphs follow power law expansion, i.e., $\{H_{\le d_{1}},$ $\ldots,$ $H_{\le d_{i}}\}$ equals $\{|H_{\le d_{1}}|\Delta^{0}, $ $\ldots,$ $|H_{\le d_{1}}|\Delta^{i-1} \}$.
\end{property}

The two properties help us introduce and prove a lemma as follows.

\begin{lemma}\label{le:ulbound}
    Let $H_{\le d_{i-1}}$, $H_{\le d_{i}}$ be the last two $d$ radius subgraphs evaluated by the expanding stage, we have $|E(H_{\leq d_{i-1}})|$ $<$ $|E(H_{\leq d^{*}})|$ $<$ $|E(H_{\leq d_{i}})|$.
\end{lemma}

The correctness is clear. Firstly, when expanding, Properties~1 and 2 hold. Secondly, the expanding stage stops when $H_{d_{i}}$ is the \textit{first} $d$ radius subgraph containing a $(\rho, c)$-truss.

Next, we establish precise relationship between $|E(H_{\leq d^{*}})|$ and $|E(H_{\leq d_{i}})|$ via the lemma below.

\begin{lemma}\label{le:prebound}
     Let $H_{\le d_{i}}$ be the last $d$ radius subgraph evaluated by the expanding stage, the inequality $\frac{|E(H_{\leq d_{i}})|}{|E(H_{\leq d^{*}})|}$ $<$ $\Delta$ holds.
\end{lemma}


Now, let us show the tight bound that is guaranteed by applying the proposed expanding strategy.

\begin{lemma}\label{le:sumsize}
     Let $(H_{\le d_{1}}$, $\ldots,$ $H_{\le d_{i}})$ be the set of $d$ radius subgraphs evaluated in order by the expanding stage, the inequality $\sum_{j=1}^{i}|E(H_{\le d_{j}})|$ $\le$  $(1+\frac{\Delta}{\Delta-1})|E(H_{\le d_{i}})|$ must hold.
\end{lemma}

\noindent\textbf{Proof sketch}. Since we have $\Delta$ invariant, $\sum_{i=1}^{i}E(H_{\le d_{j}})$ is essentially the sum of a geometric progression with a common ratio of $\frac{1}{\Delta}$ and a scale factor of $|E(H_{\le d_{i}})| $. As such it equals to $\frac{1-(\frac{1}{\Delta})^{i}}{1-\frac{1}{\Delta}}|E(H_{\le d_{i}})|$ and is no greater than $\frac{1}{1-\frac{1}{\Delta}}|E(H_{\le d_{i}})|$, which can be expressed as $(1+\frac{\Delta}{\Delta-1})|E(H_{\le d_{i}})|$. $\qed$

\noindent\textbf{Discussion}. With Lemma 3, the correctness of the following statement is clear. The running time of lines 3 to 8 in Algorithm~\ref{alg:eralg} is proportional to $(1+\Delta^{a} + \frac{1}{\Delta^{a}-1})$ $\times$ the time complexity of \textsc{ispcTrussIn($H_{\le d^{*}}$}), where $a$ is determined by the time complexity of \textsc{ispcTrussIn($H_{\le d^{*}}$}) (later on we show $a$ equals $1.5$).
This provides a tight bound for the expanding stage if we can access every $H_{\le d}$ locally during the loop of lines 3 to 8.
As such, we will introduce techniques that ensure local explanation during the expanding stage.




\noindent\textbf{Local exploration}. We propose a structure aiding us to retrieve $H_{\le d}$ for some $d$ radius subgraph with time liner to $|E(H_{\le d})|$.
We firstly show lemma as follows.

\begin{lemma}~\label{le:A}
    For any maximal connected $(\rho, c)$-truss $H$ and fixed query, there is a structure that takes $\mathcal{O}(|E(H)|)$ space, that can be built in $\mathcal{O}(|V(H)|\log(|V(H)|))$ time, and that retrieves $E(H_{\le d})$ in $\mathcal{O}(|E(H_{\le d})|)$ time.
\end{lemma}

\noindent\textbf{The structure}. The structure is an array of edges in $E(H)$ with non-decreasing order according to their distances to query location, where the distance from $\lambda$ to an edge $(u,v)$ is measured the same as Definition~\ref{def:spatialPro}.
To create the structure, we firstly sort the vertices in $H$ taking $\mathcal{O}$ $(|V(H)|\log_{2}(|V(H)|))$.
And then arrange edges into appropriate position.
For different maximal connected  $(\rho, c)$-truss $H$, we sort them separately and then merge together to speed up the performance.

With the structure, for consecutive evaluated $d$ and $d'$, we can easily retrieve $H_{\le d'}$ based on $H_{\le d}$  with time liner to $|E(H_{\le d'})\setminus E(H_{\le d})|$.

\subsection{Initial Expanding Range}
Intuitively, if the initial search range is close to $d^{*}$, the total amount of subgraphs that has to be evaluated to approaching $H_{\le d^{*}}$ is less.
This motivates us to study a lower bound of $d$ radius subgraph.

We define the lower bound $d$ radius subgraph, denoted as $H_{\le \underline{d}}$ defined as follows.

\begin{definition}\textbf{$H_{\le \underline{d}}$}.
A subgraph $H_{\le \underline{d}}$ of $H$ is a lower bound $d$ radius subgraph of $H_{\le d^{*}}$ if it satisfies conditions: 1) $H_{\le \underline{d}}$  is connected, 2) $H_{\le \underline{d}}$  satisfies keyword vertex constraint and 3) there is no $H'\subseteq H_{\le \underline{d}}$ such that $H'$ satisfies the first two constraints and $dist(\lambda,H^{\prime})$ $<$ $dist(\lambda,$ $H_{\le \underline{d}})$.
\end{definition}

$H_{\le \underline{d}}$ relaxes the structure constraint of \spatialG. As such, it can be computed efficiently, discussed below.

\begin{algorithm}[t]
\scriptsize
\KwIn{$H$}
\KwOut{$H_{\le \underline{d}}$}
\tcc{W.o.l.g, $DIST(\lambda,u)$ $\le$ $DIST(\lambda,v)$}
\ForEach{$(u,v)$ $\in$ sorted edge list of $H$}{
    maintain adjacency list\;
    initialise set rooted as $u$ and $v$ if necessary\;
    $ru$ $\leftarrow$ find(u), $rv$ $\leftarrow$ find(v)\;
    \tcc{W.o.l.g, $ru.rank$ $\le$ $rv.rank$}
    \If{$ru$ $\ne$ $rv$}{
        standard union opertion\;
        $flag$ $\leftarrow$ $true$\;
        \ForEach{$k$ $\in$ $\qk$}{
            $ru. k$ $\leftarrow$ $ru.k$ $+$ $rv.k$\;
            \If{$ru. k$ $<$ $\rho$}{
                $flag$ $\leftarrow$ $false$\;
            }
        }
       \If{flag}{
            $H_{\le \underline{d}}$ $\leftarrow$ maintained adjacency list\;
            return $H_{\le \underline{d}}$ \;
       }
    }
}

\caption{Finding lower bound search range}\label{alg:lowerbound}
\end{algorithm}

\noindent\textbf{Finding lower bound $d$ radius subgraph}.
Algorithm~\ref{alg:lowerbound} demonstrates the major steps for finding $H_{\le \underline{d}}$.
It is a refined union-find process~\cite{Tarjan:1975:EGB:321879.321884}.
We augment the union-find data structure with keyword vertex frequency.
Algorithm~\ref{alg:lowerbound} progressively performs union operations on edges in non-increasing order of their distance to $\lambda$.
By union operations, vertices that are connected are added into the same set.
Each set is attached with keyword vertex frequency for each keyword.
When an edge $(u,v)$ is being evaluated, Algorithm~\ref{alg:lowerbound} first finds if $u$ and $v$ are contained in the same set in existing union-find structure (lines 3 to 5).
If not, the two sets containing $u$ and $v$ shall be connected via standard union operation and keyword vertex frequency of the two sets shall be aggregated (lines 8 to 9).
Due to the space limitation, the discussion for union-find operations is omitted.
After a union operation, if there is a set satisfying keyword vertex constraint, we find $H_{\le \underline{d}}$. Otherwise, Algorithm~\ref{alg:lowerbound} continues.

\noindent\textbf{Time complexity}. The time complexity of Algorithm~\ref{alg:lowerbound} is $\mathcal{O}(\alpha (|V(H_{\le d^{*}})|)|E(H_{\le d^{*}})|)$, where $\alpha (|V(H_{\le d^{*}})|) \le 5$ is the cost of one union-find operation~\cite{Tarjan:1975:EGB:321879.321884} and  there are at most $|E(H_{\le d^{*}})|$ number union-find operations.
Additionally, checking keyword vertex constraint can be considered taking constant time assuming $|\qk|$ is small.


\noindent\textbf{Example}. Figure~\ref{fig:findinglowerbound} shows the keyword-aware union-find structure maintained by Algorithm~\ref{alg:lowerbound} for the query in Example~\ref{ex:mv}. Each of the sets in terms of trees in the keyword-aware union-find structure indicates a connected component the current subgraphs. After $(f,h)$ is added, the tree rooted by $h$ becomes the first connected component satisfying the keyword vertex constraint. The induced subgraphs of vertices in the trees are displayed in Figure~\ref{fig:findinglowerbound}(b).

\noindent\textbf{Alternative initial bound}. We may also relax the keyword vertex constraint to derive an alternative bound, i.e., considering the smallest $H_{\le d}$ containing a connected $c$-truss as a lower bound. But, this bound is costly to compute. 

\subsection{Checking ($\rho$, $c$)-truss in $d$ Radius Subgraph}
In this section, we show the detailed implementation of checking $(\rho,c)$-truss in a $d$ radius subgraph $H_{\le d}$, i.e., the procedure $\textsc{isptTruss}$ in Algorithm~\ref{alg:eralg}.

To simplify the discussion, for any two consecutive $H_{\le d}$ and $H_{\le d'}$ with $\frac{|H_{\le d'}|}{|H_{\le d}|} = \Delta$, let us introduce a new notation $H_{d'\setminus d}$ to denote the subgraph of $H_{\le d'}$ induced by vertices appearing in edges of $E(H_{\le d'})$ $\setminus$ $E(H_{\le d})$.

\noindent\textbf{Baseline approaches}. For checking whether there is any $(\rho,c)$-truss in $H_{\le d}$, one baseline approach is to compute the trussness for the entire $H_{\le d}$, and traverse $c$ truss subgraphs to further verify keyword vertex constraint and connectivity.
A better approach is for any two consecutive $H_{\le d}$ and $H_{\le d'}$, we update trussness for $H_{\le d}$ according to the difference between $H_{\le d'}$ and $H_{\le d}$ and traverse the updated $c$-truss for checking keyword vertex constraint and connectivity.

The two baseline approaches suffer from two drawbacks. Firstly, trussness for the whole $H_{\le d}$ is computed/updated. As such for the parts of $H_{\le d}$ that cannot contain \spatialG, the truss computation is wasted.
Secondly, checking keyword vertex constraint and connectivity has to traverse the whole $H_{\le d}$. If we can perform the check incrementally, the performance can be improved.
We propose novel techniques to address the two drawbacks.



To address the first drawback, we propose lazy $(\rho, c)$-truss checking strategy as follows.

\begin{figure}[t]
\vspace{-10pt}
  \centering
  	\subfloat[]{\includegraphics[height=30mm]{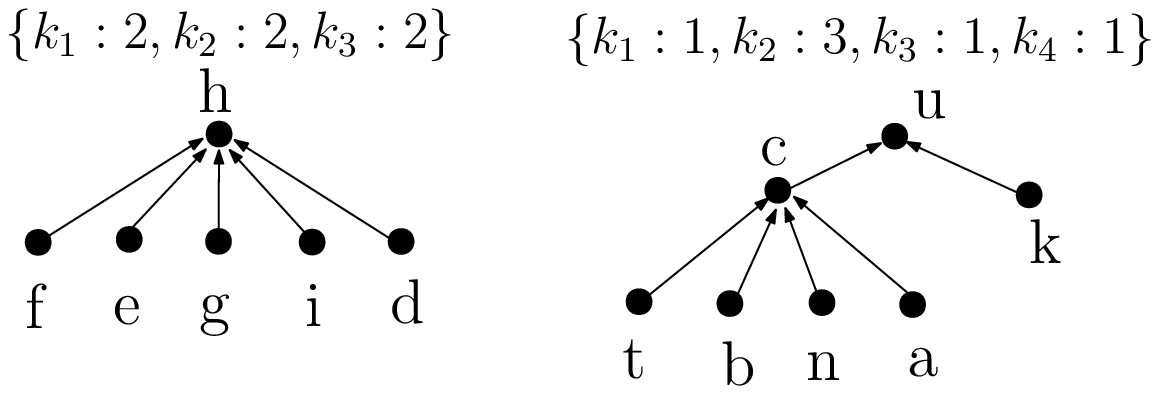}}
	\hspace{-0.2cm}
	\subfloat[]{\includegraphics[height=25mm]{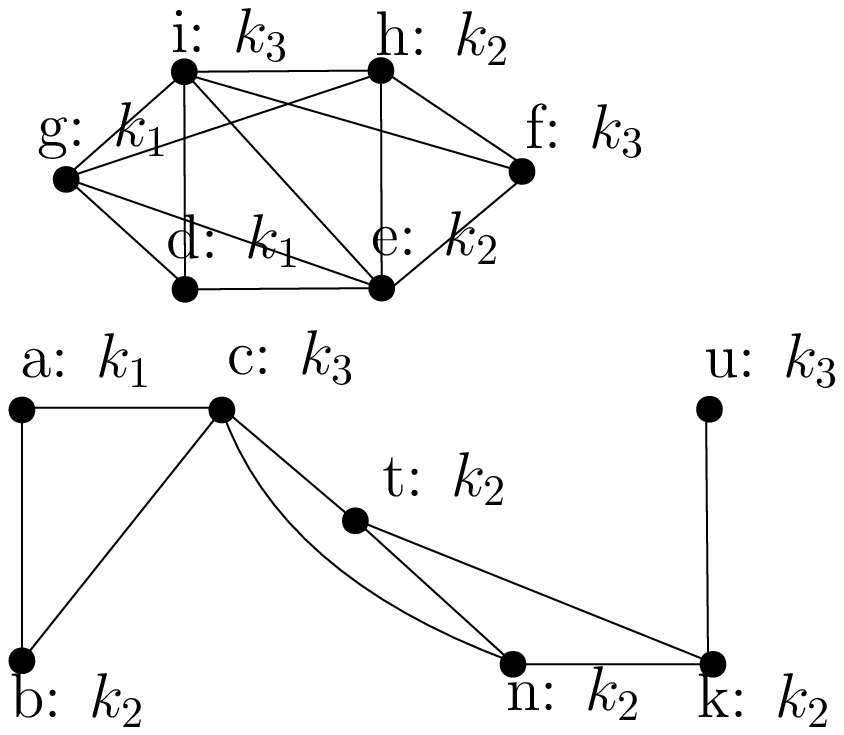}}	
  \caption{Finding $H_{\le \underline{d}}$}\label{fig:findinglowerbound}
  \vspace{-8pt}
\end{figure}

\noindent\textbf{Lazy $(\rho, c)$-truss checking strategy}. Given $H_{\le d}$, we only apply $(\rho, c)$-truss checking on any subgraph potentially containing $(\rho, c)$-truss, defined as $\rho$ potential subgraph below.

\noindent\textit{\underline{$\rho$ potential subgraph $P_{\le d}$}}. A subgraph $P_{\le d}$ $\subseteq$ $H_{\le d}$ is defined as $\rho$ potential subgraph if it is connected, satisfies keyword vertex constraint and is \textit{maximal} within $H_{\le d}$.

\noindent\textit{\underline{The strategy}}. Since a $(\rho, c)$-truss should reside in $P_{\le d}$, we propose lazy $(\rho, c)$-truss checking strategy that applies $(\rho, c)$-truss constraint checking on every $P_{\le d}$ in $H_{\le d}$ only instead of the entire $H_{\le d}$.

Identifying all $P_{\le d}$ can be done almost at no cost by using keyword aware union-find structure discussed in Algorithm~\ref{alg:lowerbound}. That is, when expanding $H_{\le d}$ to $H_{\le d'}$, vertices in edges of $H_{\le d'}$ are progressively added to keyword aware union-find structure.
As such, the $\rho$ potential subgraphs in $H_{\le d'}$ can be retrieved easily since every set in keyword aware union-find structure satisfying keyword vertex constraint identifies a $\rho$ potential subgraph.

For instance, in Figure~\ref{fig:findinglowerbound}, after all edges in $H_{\le \underline{d}}$ are retrieved, the $\rho$ potential subgraph for the query in Example~\ref{ex:mv} is $\{h,$ $f,$ $e,$ $g,$ $i,$ $d \}$ induced subgraph.
As such, according to lazy $(\rho, c)$-truss checking strategy, we only apply $(\rho, c)$-truss checking on this potential subgraph.
In contrast, we will not apply $(\rho, c)$-truss checking on subgraph induced by $\{a, b, c, t, h, k, u\}$.


Next, we show how to address the second drawback.
Please be noted, the computation discussed below shall be performed on $\rho$ potential subgraphs only. The size of these subgraphs is vastly restricted compared to the size of $H_{\le d}$.

\noindent\textbf{Union with existing truss}. To avoid graph traversing for checking keyword vertex constraint and connectivity after updating trussness, we propose a solution below.
Firstly, we maintain every maximal connected $c$ truss subgraph in every $P_{\le d}$, each of which is attached with keyword vertex frequency.
Secondly, after $P_{\le d}$ is expanded to $P_{\le d'}$, we update the maintained $c$-truss subgraphs if applicable.
Although this approach cannot update trussness for existing truss subgraphs precisely, it is sufficient and efficient to check the existence of $(\rho, c)$-truss in $P_{\le d'}$.
As such, keyword vertex constraint and connectivity checking for truss subgraphs can be performed simultaneously and incrementally.
We give formal explanations below and focus on truss unions for expanding a $P_{\le d}$ to $P_{\le d'}$. Since all $P_{\le d}$ in $H_{\le d}$ are disjoint, the truss union for expanding a $P_{\le d}$ to $P_{\le d'}$ can be easily extended to truss unions for expanding $H_{\le d}$ to $H_{\le d'}$.

\noindent\textit{\underline{Existing truss $\mathcal{C}_{\le d}$}}. We maintain connected $c$-truss subgraphs $\mathcal{C}_{\le d}\subseteq P_{\le d}$ if they exist. For each $C_{\le d}\in \mathcal{C}_{\le d}$, its keyword vertex frequencies for every keyword in $\qk$ are recorded.

\noindent\textit{\underline{Truss potential subgraph}}. After expanding $P_{\le d}$ to $P_{\le d'}$. We only compute maximal truss subgraphs in \textit{truss potential subgraph} defined below.

\begin{definition}\label{def:tps} \textbf{Truss potential subgraph.}
Given two consecutive $P_{\le d}\subseteq H_{\le d}$ and $P_{\le d'}\subseteq H_{\le d'}$ with $\mathcal{C}_{\le d}\subseteq P_{\le d}$, the truss potential subgraph is defined as $TP_{d\setminus d'} = H_{\le d'}(V')$, where $V'$ is the set of vertices appearing in $E(P_{\le d'})$ $\setminus$ $E(\mathcal{C}_{\le d})$.
\end{definition}

The sufficiency of $TP_{d\setminus d'}$ is clear since it contains all triangles in $P_{\le d'}$ for edges that are not in $\mathcal{C}_{\le d}$ but potentially lead to $(\rho,c)$-truss.


\noindent\textit{\underline{Truss union}}. Based on Definition~\ref{def:tps}, for consecutive $P_{\le d}$ and $P_{\le d'}$, we compute maximal truss subgraphs in $TP_{d\setminus d'}$ and then add them to $\mathcal{C}_{\le d}$ via union operation, which forms $\mathcal{C}_{\le d'}$.

\begin{figure}
  \centering
  	\subfloat[]{\includegraphics[height=30mm]{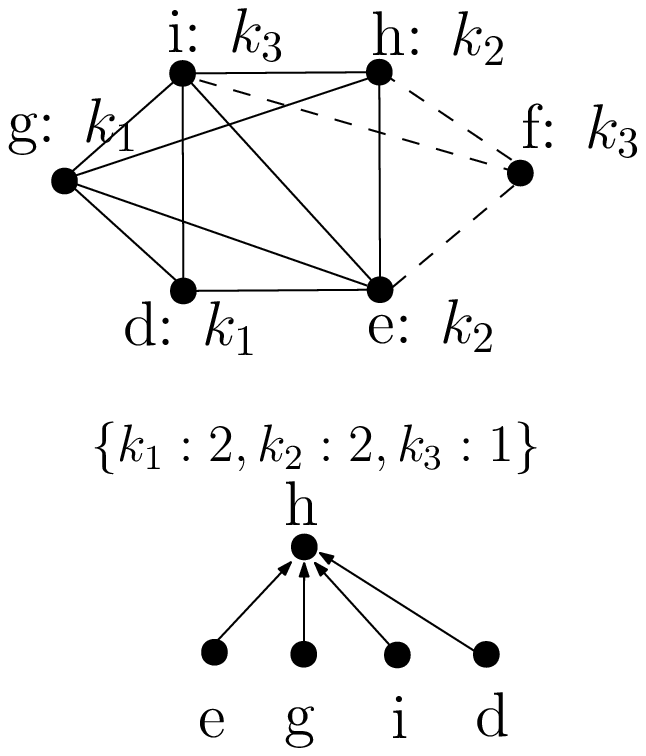}}
	\hspace{-0.1cm}
	\subfloat[]{\includegraphics[height=30mm]{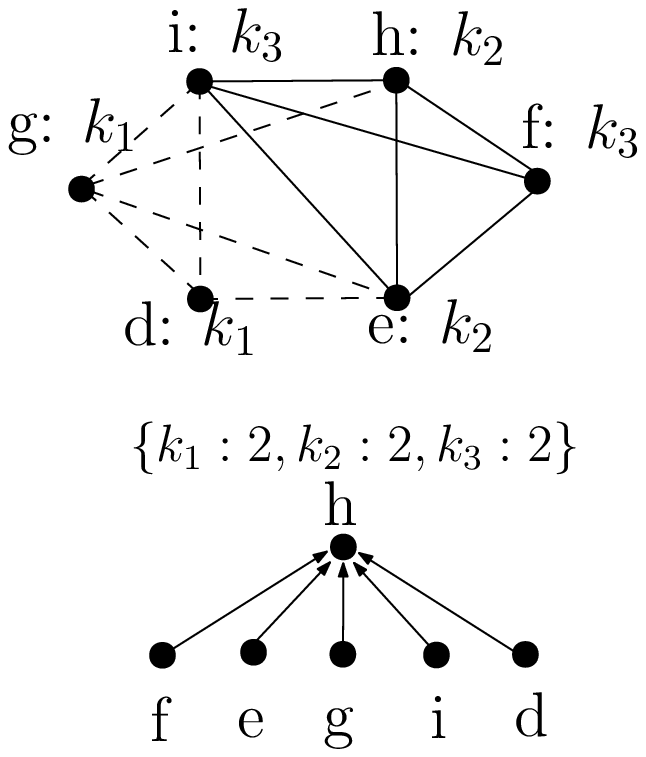}}	
  \caption{Truss union}\label{fig:truusunion}
\end{figure}

\noindent\textbf{Example}.  In Figure~\ref{fig:truusunion}, we show an example for truss union operation. In Figure~\ref{fig:truusunion}(a), let $\{g,$ $d,$ $e,$ $i,$ $h\}$ induced subgraph be $H_{\le d}$ and its $P_{\le d}$ and  $\mathcal{C}_{\le d}$ are the same graph. Let the whole graph in Figure~\ref{fig:truusunion}(a) be $H_{\le d'}$. Then, $P_{\le d'}$ is $\{g,$ $d,$ $e,$ $i,$ $h,$ $f\}$ induced subgraph, and $E(P_{d'})$ $\setminus$ $E(\mathcal{C}_{\le d})$ is $\{$ $(f, e),$ $(f, h)$, $(f,i)$ $\}$. Then $TP_{d'\setminus d}$ is $\{i,$ $h,$ $f,$ $e\}$ induced subgraph shown in Figure~\ref{fig:truusunion}(b). Since there is a $c$-truss in $\{i,$ $h,$ $f,$ $e\}$ induced subgraph, truss-union data structure in Figure~\ref{fig:truusunion}(a) (in terms of tree structure) is updated to the one in Figure~\ref{fig:truusunion}(b).


Next, we show the ($\rho$,$c$)-truss checking algorithm with the proposed techniques.

\noindent\textbf{The ($\rho$,$c$)-truss checking algorithm}. The principal steps of ($\rho$,$c$)-truss checking are shown in Algorithm~\ref{alg:rttrusschecking}.

\noindent\textit{Data structure}. Since Algorithm~\ref{alg:rttrusschecking} is called iteratively, it works on progressively refined data structures including adjacency list of $H_{\le d}$, the keyword aware union-find denoted as $UF_{\le d}$ storing every $\rho$ potential subgraph, the keyword-aware truss union-find structure denoted $TUF_{\le d}$ storing maximal connected $k$-truss with aggregated keyword frequency. All those data structures are empty sets before the first time when Algorithm~\ref{alg:rttrusschecking} is called.

\noindent\textit{Principal steps}. Algorithm~\ref{alg:rttrusschecking} adds each edge in $H_{d'\setminus d}$ to $H_{\le d'}$ and performs union operation on each edge to $UF_{\le d'}$, where the edges of $H_{d'\setminus d}$ can be retrieved easily with the sorted array proposed in Lemma~\ref{le:A}.
After that, Algorithm~\ref{alg:rttrusschecking} computes maximal $c$-truss subgraphs in the truss potential subgraph defined in Definition~\ref{def:tps} (line 5).
More precisely, with $UF_{\le d'}$ and $TUF_{\le d'}$, $TP_{d'\setminus d}$ is $H_{\le d'} (V')$, where $V'$ are the vertices appearing in $E($ $\cup_{P_{\le d'} \in  UF_{\le d'}}$ $P_{\le d'})$ $\setminus$ $E(\cup_{\mathcal{C}_{\le d}\in TUF_{\le d'}}\mathcal{C}_{\le d})$.
Next, Algorithm \ref{alg:rttrusschecking} performs truss union operations for the computed maximal $c$ truss subgraphs.
After the truss union, if there is a set in $TUF_{\le d'}$ that satisfies keyword vertex constraint, then there is a $(\rho, c)$-truss and Algorithm~\ref{alg:rttrusschecking} returns the $(\rho, c)$-truss $S$ (line 9).
Otherwise, Algorithm~\ref{alg:rttrusschecking} returns $\emptyset$.

The correctness of Algorithm~\ref{alg:rttrusschecking} is clear according to the techniques discussed above.

\noindent\textbf{Time complexity}. The time complexity of Algorithm~\ref{alg:rttrusschecking} is $\mathcal{O}(|E(H_{\le d'})|^{1.5})$. Computations between lines $2$ to $4$ are dominated by keyword aware union-find operations that are $\mathcal{O}(|E(H_{d\setminus d'})|)$, and it is the same for lines 6 to 7.
The dominating part is line 5. In the worst case, $TP_{d'\setminus d}$ could be the same as $H_{\le d'}$. This results in $\mathcal{O}(|E(H_{\le d'})|^{1.5})$ time complexity for Algorithm~\ref{alg:rttrusschecking}.

To conclude the expanding stage, we show lemma below.
\begin{lemma}
The time complexity of expanding stage is $\mathcal{O}$ $((1+\Delta^{1.5} + \frac{1}{\Delta^{1.5}-1})$ $\times$ $|E(H_{\le d^{*}}|^{1.5})$.
\end{lemma}

The correctness is clear based on the time complexity of Algorithm~\ref{alg:rttrusschecking} and Lemma~\ref{le:sumsize}. When $\Delta =2$, the time complexity becomes the minimum that is $\mathcal{O}$$(|E(H_{\le d^{*}})|^{1.5})$.
\begin{algorithm}[t]
\scriptsize
    \KwIn{$H_{\le d}, d'$}
    \KwOut{$S$}
    \tcc{$UF_{\le d}$: keyword aware union find structure storing all $\rho$ potential graphs in $H_{\le d}$}
    \tcc{$UF_{\le d}$: keyword aware truss union find structure storing all truss subgraphs in $H_{\le d}$}
    $H_{\le d'}\leftarrow$ $H_{\le d}$, $UF_{\le d'}\leftarrow UF_{\le d}$, $TUF_{\le d'}\leftarrow TUF_{\le d}$ \;
    \ForEach{$(u,v) \in H_{d'\setminus d}$ }{
        $H_{\le d'}$ $\leftarrow$ $H_{\le d'}\cup \{\{u,v\}\}$\;
        $UF_{\le d'}$ $\leftarrow$ $UF_{\le d'}$ $\cup$ $\{(u,v)\}$; \tcp{union}

    }
    \tcc{ $P_{d'\setminus d}$ has been generated during updating $UF_{\le d'}$}
    $H'$ $\leftarrow$ compute $c$-truss in $TP_{d'\setminus d}$\;
    \ForEach{$(u,v)$ $\in$ $H'$}{

        $TUF_{\le d'}$ $\leftarrow$ $TUF_{\le d'}$ $\cup$ $\{(u,v)\}$; \tcp{truss union}
    }

    \eIf{$TUF_{\le d'}$ contains a set satisfies keyword constraint}{
        \Return the set as $S$ \;

    }{
        \Return $\emptyset$\;
    }



    \caption{\textsc{incIspcTrussIn($H_{\le d}$, $d'$)}}\label{alg:rttrusschecking}
\end{algorithm} 
\vspace{-10pt}

\section{Reducing Stage}\label{sec:reducing}

For the reducing stage, we focus on searching \spatialGspace in the $(\rho,c)$-trusses found by the expanding stage, denoted as $S$. We would like to revisit that the size of $S$ is $\mathcal{O}(|H_{\le d^{*}}|)$.

Intuitively, this stage progressively removes the vertex in $S$ that is most distant to the query location till there is no $(\rho,c)$-truss in the remaining $S$.
The last survived $(\rho,c)$-truss is \spatialG.

Efficiently checking the existence of $(\rho,c)$-truss after deleting a vertex is challenging.
This is because after a vertex deletion, we have to deal with truss computation, verifying keyword vertex constraint and checking connectivity.
The obvious time consuming part is truss computation, which can be bounded nicely by taking the advantage of decremental truss computation. The pitfall when analyzing the cost is ignoring the cost of keyword vertex constraint and connectivity checking. Actually, a graph traversing-based implementation for checking them can lead to complexity of $\mathcal{O}$ $(|V(H_{\le d^{*}})|$ $|E(H_{\le d^{*}})|)$, which is worse than the time complexity of truss computation and becomes the performance \textit{bottleneck} of \spatialG~search.

We will propose efficient approach for checking multiple constraints together.

\subsection{Reducing Strategy}
In this part, we show the reducing strategy.

\noindent\textbf{The strategy}. Algorithm~\ref{alg:reducepttruss} shows the major steps of the strategy for the reducing stage. It progressively removes the vertex that is most distant to $\lambda$ (the query location) in $S$ and checks the existence of $(\rho,c)$-trusses in the remaining of $S$ after the deletion. If there exists one, Algorithm~\ref{alg:reducepttruss} continues to delete next most distant vertex in $S$.
Otherwise, Algorithm~\ref{alg:reducepttruss} returns the last $(\rho,c)$-truss as \spatialG.

Clearly, the strategy can find \spatialGspace in $S$ correctly since Algorithm~\ref{alg:reducepttruss} maintains an invariant that every time deleting the most distant vertex in $S$, $S$ contains set of $(\rho,c)$-trusses.
This invariant is ensured by our proposed \textsc{pcTrussChecking} in Algorithm~\ref{alg:reducepttruss}.
That is, after the most distant vertex is deleted, we further delete edges violating the minimum trussness requirement. Meanwhile, for each edge deletion, we immediately check whether the remaining subgraphs contain a connected subgraph satisfying keyword vertex constraint. If no, we stop edge deletions and return empty set since no $(\rho,c)$-truss exists. If yes, we exclude all the other subgraphs since they cannot lead to \spatialG.

It is clear to see that the time complexity of Algorithm~\ref{alg:reducepttruss} consists of the trussness computation cost and keyword-aware connectivity checking cost. The former is bounded by $\mathcal{O}$ $(|E(H_{\le d^{*}})|^{1.5})$ since Algorithm~\ref{alg:reducepttruss} takes the advantage of decremantal truss computation and we have shown that $S$ returned by the expanding stage will be no greater than $\mathcal{O}$ $(|E(H_{\le d^{*}})|)$. The later is dependent on the cost of \textsc{ckChecking} called in \textsc{pcTrussChecking}, Algorithm~\ref{alg:reducepttruss}.

In the following subsection, we focus on proposing techniques for devising efficient \textsc{ckChecking} (Algorithm~\ref{alg:ckChekcing}), which makes the total cost of keyword-aware connectivity checking is less than $\mathcal{O}$ $(|E(H_{\le d^{*}})|^{1.5})$.
As such, the proposed strategy embedding with elegant techniques devised by us can bound the total cost of multi-constraint checking in the reducing stage as $\mathcal{O}(|E(H_{\le d^{*}})|^{1.5})$.

\begin{algorithm}[t]
\scriptsize
\KwIn{ $S$: $(\rho, t)-truss$}
\KwOut{$S^{*}$}

	\SetKwFunction{procOne}{pcTrussChecking}

	sort vertices in $S$ according to their distance to $\lambda$ in none decreasing order\;
	
	\ForEach{$u\in V(S) $}{
		$S'$ $\leftarrow$  $\textsc{pcTrussCecking}(S, u)$\;
	
		\eIf{$S'$ $\ne$ $\emptyset$}{
				$S \leftarrow S'$; \tcp{order preserved}
		}{
				\Return $S$ as $S^{*}$\;
		}
	}
		
	\SetKwProg{iincCore}{Procedure}{}{}
	\iincCore{\procOne($S$, $u$)}{
		$Q\leftarrow \emptyset$\;
		\ForEach{$v\in $ $N(u,S)$}{
			$Q\leftarrow$ $Q \cup \{(u,v)\}$\;		
		}	
		\While{$Q$ $\ne$ $\emptyset$}{
			$(u,v)\leftarrow$ $Q.pop()$\;		
			\ForEach{$w \in N(u,S)\cap N(v,S)$}{       	
                update triangle numbers for $(w,u)$, $(w,v)$ \;
                put $(w,u)$, $(w,v)$ into $Q$ if they cannot be part of $c$-truss\;
        		remove $(u,v)$ from $S$ \;
        		\tcc{Checking connectivity and keyword constraints}
        		\If{\textsc{ckChecking($(u,v)$)}}{
        			 \Return $\emptyset$\;
        		}
        	}	
			\Return the remaining $S$\;
		}		
	}
\caption{\textsc{reducepcTruss($Q$,$S$)}}\label{alg:reducepttruss}
\end{algorithm}

\subsection{Keyword-aware Connectivity Checking}
In this section, we show how to efficiently check the existence of a connected subgraph satisfying keyword vertex constraint after an edge is deleted induced by removing the most distant vertex in Algorithm~\ref{alg:reducepttruss}.

\noindent\textbf{High level idea}. We will maintain a minimum spanning forest for $S$ (input of Algorithm~\ref{alg:reducepttruss}) augmented with aggregated keyword vertex frequency. Notice that initially, every spanning tree in the forest satisfies keyword vertex constraint.
After an edge is deleted from $S$, one of the two cases below may happen.

\noindent\textit{\underline{Case 1: the deleted edge is not in the forest}}. In this case, the remaining subgraphs are still connected and each connected subgraph still satisfies keyword vertex constraint.

\noindent\textit{\underline{Case 2: the deleted edge is in the forest}}. In this case,  one of the tree in the minimum spanning forest is cut into two trees, which may lead to one of the following subcases.

\textit{Subcase 1: cannot link the cut trees}. In this subcase, we cannot find a replacement edge from the remaining $S$ to link the two trees, which means the subgraph referred by the two trees becomes two disjoint subgraphs.
We update keyword vertex frequency for each of the cut tree. After the update, we safely prune the cut tree from the maintained spanning forest if it does not satisfy keyword constraint since they cannot contribute to \spatialG.

\textit{Subcase 2: can link the cut trees}. If we can find a replacement edge, the subgraph referred by two cut trees is still connected. We link the two trees with the replacement edge.
Keyword vertex frequency remains the same.

It is clear that the above idea can correctly maintain all connected subgraphs satisfying keyword vertex constraint if they exist after deleting an edge from $S$. But, it is challenging to preform the maintenance efficiently since checking the existence of a replacement edge could be costly.

To make the maintenance efficient, we borrow the idea from \cite{holm2001poly}.
Given $S$, every edge in $E(S)$ is associated with a level progressively increased as edges are deleted, which is equivalent to progressively partitioning $S$ hierarchically.
Edges with high level refer to a more restricted part of $S$.
In contrast, edges with low level refer to a more general part of $S$ (super graphs of the high level subgraphs).
As such when deleting an edge with a certain level, we do not need to consider any edge with lower level as a replacement edge, which elegantly reduces the search space for finding a replacement edge. 


\begin{figure}[t]
\vspace{-10pt}
	\centering
	\subfloat[]{\includegraphics[height=20mm]{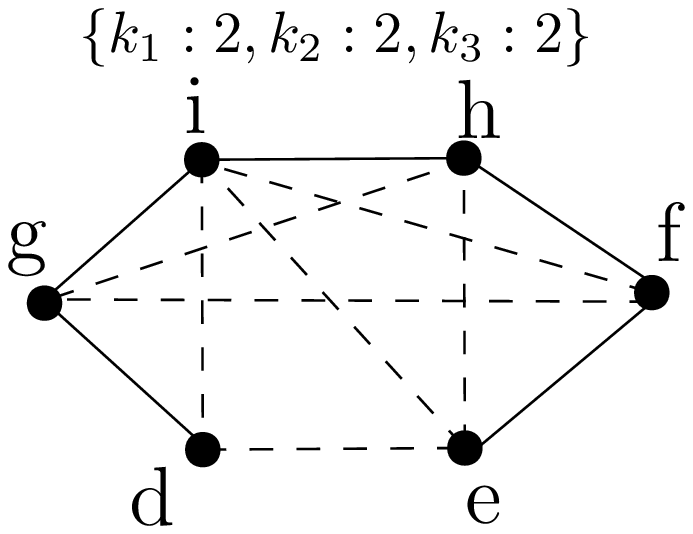}}
	\subfloat[]{\includegraphics[height=20mm]{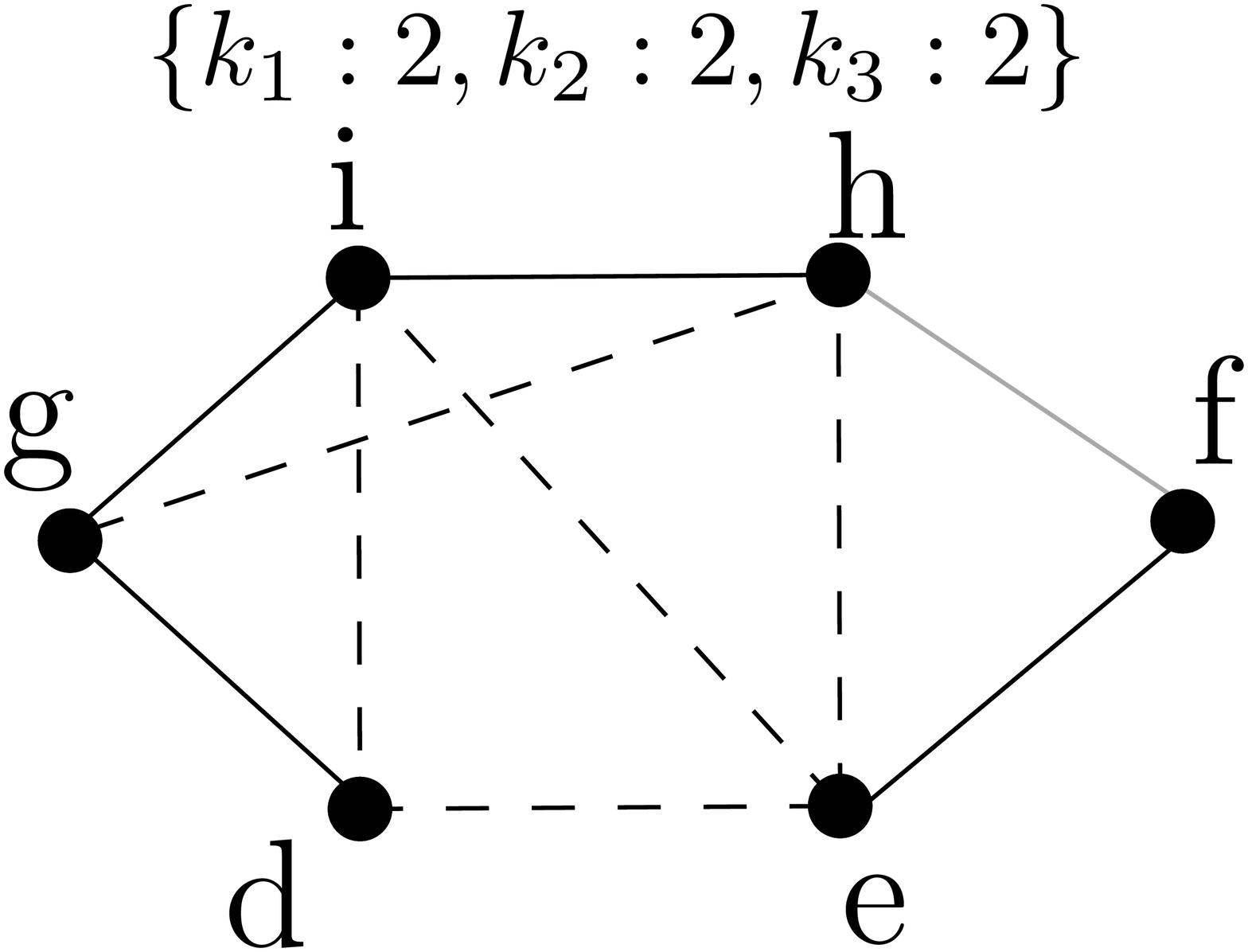}}	
    \vspace{-8pt}
    \subfloat[]{\includegraphics[height=20mm]{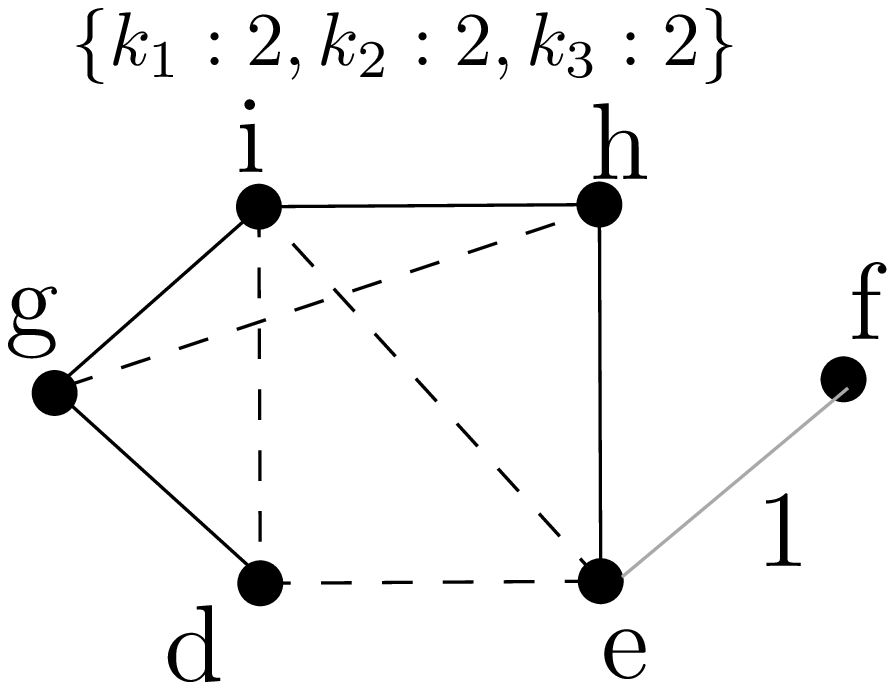}}
	\subfloat[]{\includegraphics[height=20mm]{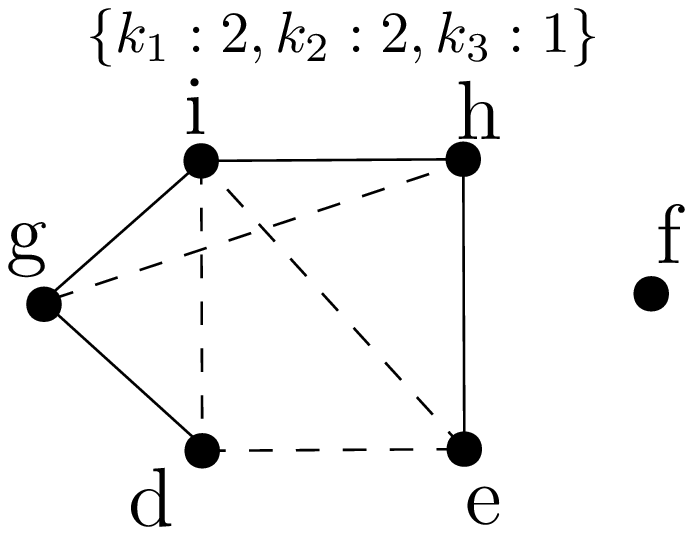}}	
	\caption{keyword aware spanning forest}\label{fig:ksf}
	\vspace{-5pt}
\end{figure}

We first use an example to demonstrate our method.

\noindent\textbf{Example}. Suppose we have the input graph as shown in Figure~\ref{fig:ksf}(a), and we want remove vertex $f$.
The minimum spanning forest is shown in Figure~\ref{fig:ksf}(a) with edges in solid lines and the edges not in the spanning forest are shown as dashed lines.
We do not show the level of an edge if its level is $0$.
Removing $f$ is equivalent to remove edges incident to $f$.
It is trivial to remove $(i,f)$ and $(g,f)$ since they are not a part of the spanning forest.
After that, supposing that we remove $(h,f)$ shown as grey line in Figure~\ref{fig:ksf}(b), the spanning tree becomes two trees where the tree with vertices of $\{f,e\}$ is the smaller tree and the level of the edge in the tree is increased by $1$.
By checking edges incident to $f$ and $e$, we find a replacement edge $(h,e)$. By connecting the two trees, the spanning tree becomes the one in Figure~\ref{fig:ksf}(c).
Next, we remove $(e,f)$ shown in  Figure~\ref{fig:ksf}(c) and the tree with vertex only $f$ is the smaller tree.
In this case, we cannot find any edge incident to $f$, leading to Figure~\ref{fig:ksf}(d).
We know the graph becomes separated and we also know that there is a connected component in the remaining graph with keyword frequencies of $\{$$k_{1}$:$2$, $k_{2}$:$2$, $k_{3}$:$1$$\}$.
Without using the proposed method, we cannot simultaneously know the keyword vertex frequency and the connectivity of the subgraph after deleting $f$.

\begin{algorithm}[t]
\scriptsize
\KwIn{$(u,v)$, $F$}
\KwOut{True or False}

    $l$ $=$ $(u,v).level$\;
    \If{$(u,v)$ $\notin$ $F_{\ge 0}$ }{
        delete $(u,v)$ directly\;
        \Return True\;
    }

    \tcc{adjust level of edges}
    $T_{u}$, $T_{v}$ $\leftarrow$ delete $(u,v)$ from $F_{\ge l}$\;
    assume $|V(T_{u})|$ $\le$ $|V(T_{u})|$\;
    \ForEach{$e\in E(T_{u})$}{
        $e.l$ $\leftarrow$ $e.l+1$\;
        progressively calculate aggregated keyword vertex frequency in $T_{u}$\;
    }

    \For{$i$ $\leftarrow$ $l+1$ to $0$}{
        cut $F_{\ge i}$ \;
    }

    \tcc{search the replacement edge of $(u,v)$}
    $altE$ $\leftarrow$ $\emptyset$\;
    \For{$i$ $\leftarrow$ $l$ to $0$}{
        \For{$v \in T_{u}$}{
            \tcc{$v.adjG_{i}$ is level-aware adjacency list}
            \For{$w\in v.adjG_{i}$}{
                \eIf{$w\in V(T_{u})$}{
                    increase level of $(v,w)$ by 1\;
                }{
                    \tcp{find the replacement edge}
                    $altE \leftarrow$ $(v,w)$, $altE.l$ $\leftarrow$ $i$\;
                    \textbf{Break} \;
                }

            }

        }
    }

    \eIf{$altE$ $\ne$ $\emptyset$}{
        \tcc{alternative edge is found and update the minimum forest}
        \For{$i$ $\leftarrow$ $altE.l$ to $0$}{
            Link corresponding two subtrees in $F_{i}$ via $altE$\;
        }

    }{
        \tcc{graph is split and keyword frequencies shall be updated}
        update aggregated keyword frequencies of $T_{v}$ in $F_{0}$ according to the aggregated keyword vertex frequencies of $T_{u}$ \;
        prune trees in $S$ do not satisfy keyword vertex constraint\;
    }

    \Return True \textbf{If} at least one of $T_{v}$ in $F_{0}$ and $T_{u}$ satisfies keyword constraint \textbf{else} \Return False\;

\caption{\textsc{ckChekcing}($(u,v)$)}\label{alg:ckChekcing}
\end{algorithm}

Now, let us describe the method formally. We first introduce the keyword aware spanning forest.

\noindent\textbf{Keyword aware spanning forest}. The minimum spanning tree for every connected ($\rho$, $c$)-truss in $S$ from the expanding stage is computed and stored, in which each spanning tree is augmented with keyword vertex frequency. As discussed, initially every spanning tree in this forest ($F$) satisfies keyword vertex constraint and level for every edge in $S$ is assigned as $0$. Below, we use $F_{\ge i}$ to denote the forest of edges with level at least $i$.

\noindent\textbf{The algorithm}. Algorithm~\ref{alg:ckChekcing} guarantees that after an edge deletion, every remaining minimum spanning tree in the keyword aware spanning forest satisfies keyword vertex constraint. It returns true if the keyword aware spanning forest is not an empty set. Otherwise it return $\emptyset$. To efficiently achieve that, Algorithm~\ref{alg:ckChekcing} maintains invariants as follows.

\textit{Invariant 1}. $F_{\ge 0}\supseteq F_{\ge 1}\supseteq,\ldots, \supseteq F_{\ge lmax}$ always holds. This invariant ensures no duplicated trees are generated.

\textit{Invariant 2}. $F_{\ge i}$ is a minimum spanning forest for edges with level at least $i$ induced subgraphs. This invariant maximizes the possibility that a deleted edge is not in the maintained forest.

\textit{Invariant 3}. The number of vertices in $F_{\ge l}$ is always no greater than $\lfloor \frac{|V(S)|}{2^{l}}\rfloor$. This is because when a tree is split into to subtrees,  Algorithm~\ref{alg:ckChekcing} always increases the levels of edges in the smaller tree by $1$. As such, the worst case is that every time a tree is split, the two trees are equal size, leading to largest possible size of a tree at level $l$ as $\lfloor \frac{|V(S)|}{2^{l}}\rfloor$.
This invariant guarantees that the level of an edge is no greater than $\log_{2}|V(S)|$, which is the key for time complexity analysis.

More detailed steps are given below.

Given that $(u,v)$ with level $l$ is to be deleted, Algorithm~\ref{alg:ckChekcing} firstly checks whether it is in the current forest or not.

\noindent\textit{\underline{Case 1: }}. If $(u,v)$ is not in $F_{\ge 0}$, $(u,v)$ is deleted (line 3), the algorithm return true .

\noindent\textit{\underline{Case 2: }}. If $(u,v)$ is in $F_{\ge 0}$, Algorithm~\ref{alg:ckChekcing} deletes it from the tree containing $(u,v)$ from level $l$ which is the highest forest it is in.

\textit{Performing tree cut} (lines 10 to 11). The tree is cut into two subtrees $T_{u}$ and $T_{v}$, and levels of edges in the smaller tree in terms of number of vertices are increased by $1$.
Next, Algorithm~\ref{alg:ckChekcing} propagates the deletion from $F_{\ge l+1}$ to $F_{\ge 0}$ so that from the view at all the levels, the tree is split.

\textit{Searching a replacement edge} (line 13 to 20).  After performing tree cut, Algorithm~\ref{alg:ckChekcing} starts to search a replacement edge of $(u,v)$ that may connect $T_{u}$ to $T_{v}$. This is achieved by searching all edges incident to vertices appearing in $T_{u}$.
To maintain the minimum spanning forest property, Algorithm~\ref{alg:ckChekcing} searches an alternative edge from level $l$.
If an edge $(v,w)$ is incident to $T_{u}$ but $v$ and $w$ are in $T_{u}$ then its level is increased by $1$.

\textit{Subcase 1: cannot link the cut trees} (lines 24 to 26). If no replacement edge is found, $T_{u}$ induced subgraph is isolated. Aggregated keyword frequencies are adjusted. If the there is a tree violating the keyword vertex constraint after the adjustment, it is pruned.

\textit{Subcase 2: can link the cut tress} (lines 21 to 23). If a replacement edge is found, the incident edge $(v,w)$ shall link $T_{u}$ to $T_{v}$ and this edge is inserted to $F_{\ge l}$ to $F_{\ge 0}$ so that from the view of all the levels, the tree is linked.

Next we further discuss data structures used in our implementation which are useful for time complexity analysis.

\noindent\textbf{Data structure}. In Algorithm~\ref{alg:ckChekcing}, to efficiently deal with tree cut and tree link operations, we store spanning forest as Euler tours and the Euler tours are stored as balanced binary search tree~\cite{henzinger1999randomized}.
As such, each operation of tree cut and tree link can be performed in $\mathcal{O}(\log_{2}(|V(S)|))$.

\noindent\textbf{Time complexity analysis}.
The time complexity of index initialisation is  $\mathcal{O}(|E(S)| log_{2}(|V(S)|))$.
The time complexity of Algorithm~\ref{alg:ckChekcing} for deleting $E(S)$ number of edges is $\mathcal{O}(|E(S)| log_{2}^{2}(|V(S)|))$.
Lines 7 to 9 in the algorithm have the time complexity of $\mathcal{O}(|E(S)| log_{2}(|V(S)|))$.
This is because for each edge, its level is at most $log_{2}(|V(S)|)$.
The dominating parts are lines 14 to 23 and lines 24 to 27 in the algorithm since they perform up to $\mathcal{O}(log_{2}(|V(S)|))$ number of cut or link operations and each has a cost of $\mathcal{O}(\log_{2}(|V(S)|))$, which results $\mathcal{O}(|E(S)| log_{2}^{2}(|V(S)|))$.

Due to the space limitation, the discussion of obvious prunings is omitted.

\vspace{-10pt}

\subsection{Search Algorithm Wrap-Up}\label{sec:warp}
We formally claim the lemma as follows.

\begin{lemma}
   The time complexity of \spatialG~is the maximum of $\mathcal{O}$ $(|E(H_{\le d^{*}})|^{1.5})$ and $\mathcal{O}(|V(H)|\log_{2}|V(H)|)$.
\end{lemma}

The correctness is clear given the discussion throughout this paper.
In practical, our proposed algorithm is much faster since we propose many optimizations that prune search spaces as much as possible.
We shall evaluate those optimisations in experimental studies.

Below, we introduce some of other possible constraints that can be solved efficiently by the proposed search framework and then establish the lower bound for the proposed framework.

\noindent\textbf{Alliterative keyword constraints}. We can use Jaccard similarity to measure the keyword similarity between the attributes of a vertex and the query keywords firstly and then set minimum similarity threshold as the keyword constraint for the desired geo-social group.

\noindent\textbf{Alliterative size constraints}.  We can directly set a minimum size as the size constraint for a geo-social group. Or we set the minimum vertex frequency for each of the keyword vertex to define the size constraint.

\noindent\textbf{Alliterative social constraints}. Our proposed method supports social constraint defined as $k$-core, or more generalized cohesive constraint $(k,s)$-nucleus.

\noindent\textbf{Influential constraints}. Beside keyword, size and social constraints, we can consider member influence as a factor for finding the geo-social group. Assuming each vertex in a group has an influential score, we set minimum influential score threshold as the influential constraint for the desired geo-social group.

Given multiple polynomial checkable constraints, let us use $\mathcal{O}(\mathcal{C}_{max})$ to denote the dominating time complexity for checking all the constraints.
We are ready to establish a general lower bound for geo-social group search problem with multiple constraints using the proposed search framework.

\begin{lemma}
   The lower bound of multi-constraint geo-social group search is $\Omega(\mathcal{C}_{max})$ using the proposed search framework.
\end{lemma}

For the instance of multi-constraint geo-social group search, \spatialG~search, our proposed techniques ensure that the time complexity of the search matches this lower bound.

\begin{table}
\centering
\scriptsize
	\begin{tabular}{|c|c|c|}
		\hline
		Parameter	&Range	&Default value	\\
		\hline
        $c$ & $3$, $4$, $5$, $6$, $7$, $8$  & $6$   \\	
        \hline
		$|\qk|$ &	$1$, $3$, $5$, $7$, $9$	 & $3$\\	
		\hline
		$\rho$ &	$1$, $3$, $5$, $7$, $9$	 & $3$ \\	
		\hline
	\end{tabular}
	\caption{Parameter settings}\label{tab:para}
\end{table}

\begin{table}
\scriptsize
\centering
	\begin{tabular}{|r|r|r|r|r|}
		\hline
		Dataset & \#vertices & \#edges & \#checkins & $c_{max}$\\
		\hline
		Gowalla & 196,591 & 950,327 & 6,442,890 & 29\\
		\hline
		Brightkite & 58,228 & 214,078 & 4,491,143 & 43\\
		\hline
		Foursquare & 4,899,219 & 28,484,755 &	1,021,970 & 16\\
		\hline
		Weibo & 1,019,055 & 32,981,833 & 32,981,833 & 11\\
		\hline
		Yelp & 257,532 &  957,711 & 431,563 & 21\\
        \hline
        WoW$$    & 278  &  752 & 278   & 36 \\
        \hline
		\end{tabular}
	\caption{Statistic information in datasets}~\label{tab:dataset}
	\vspace{-10pt}
\end{table}

\begin{figure*}[ht]
\vspace{-10pt}
	\centering

    \subfloat[Gowalla]{\includegraphics[width=3.7cm]{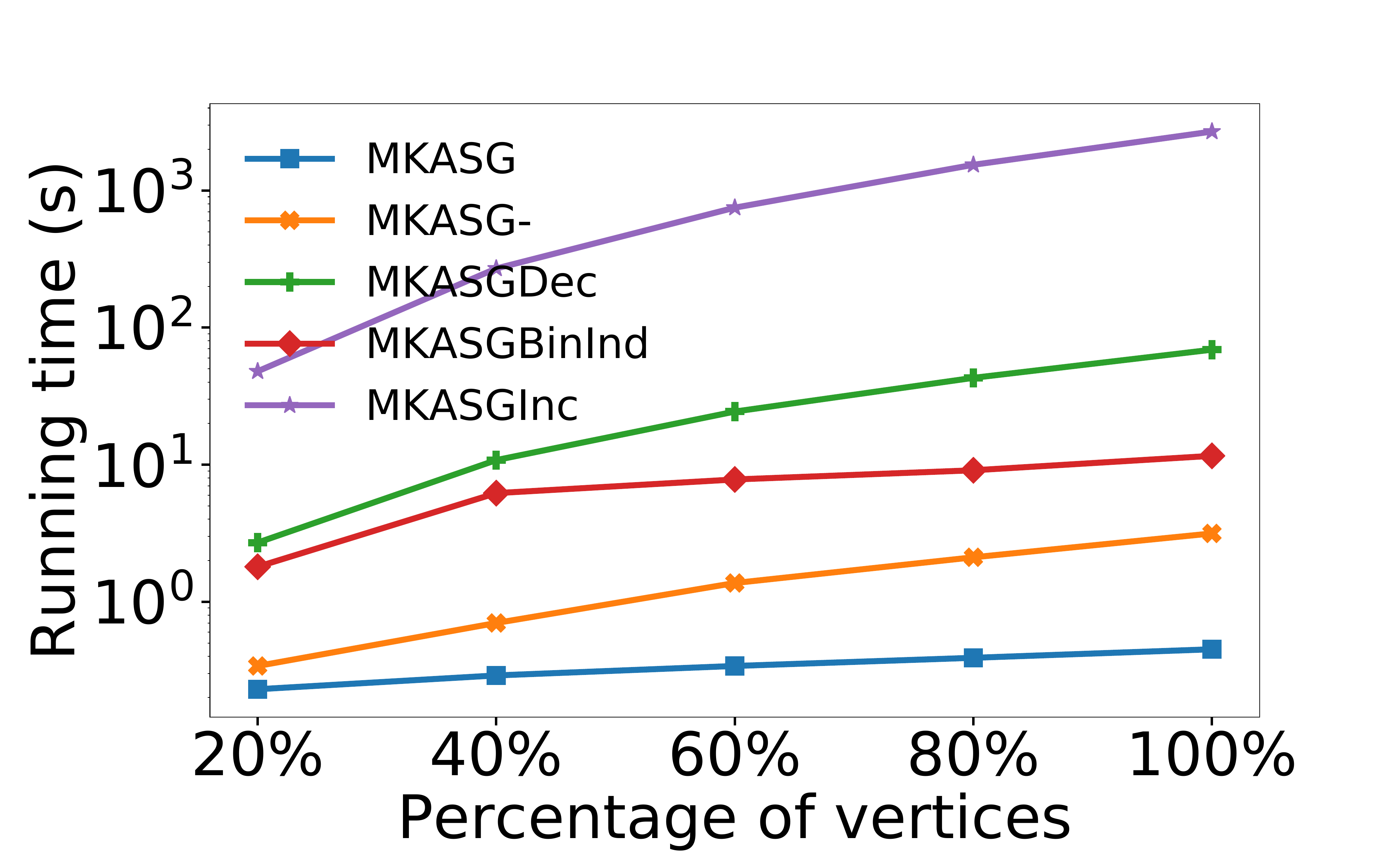}}
	\subfloat[Brightkite]{\includegraphics[width=3.7cm]{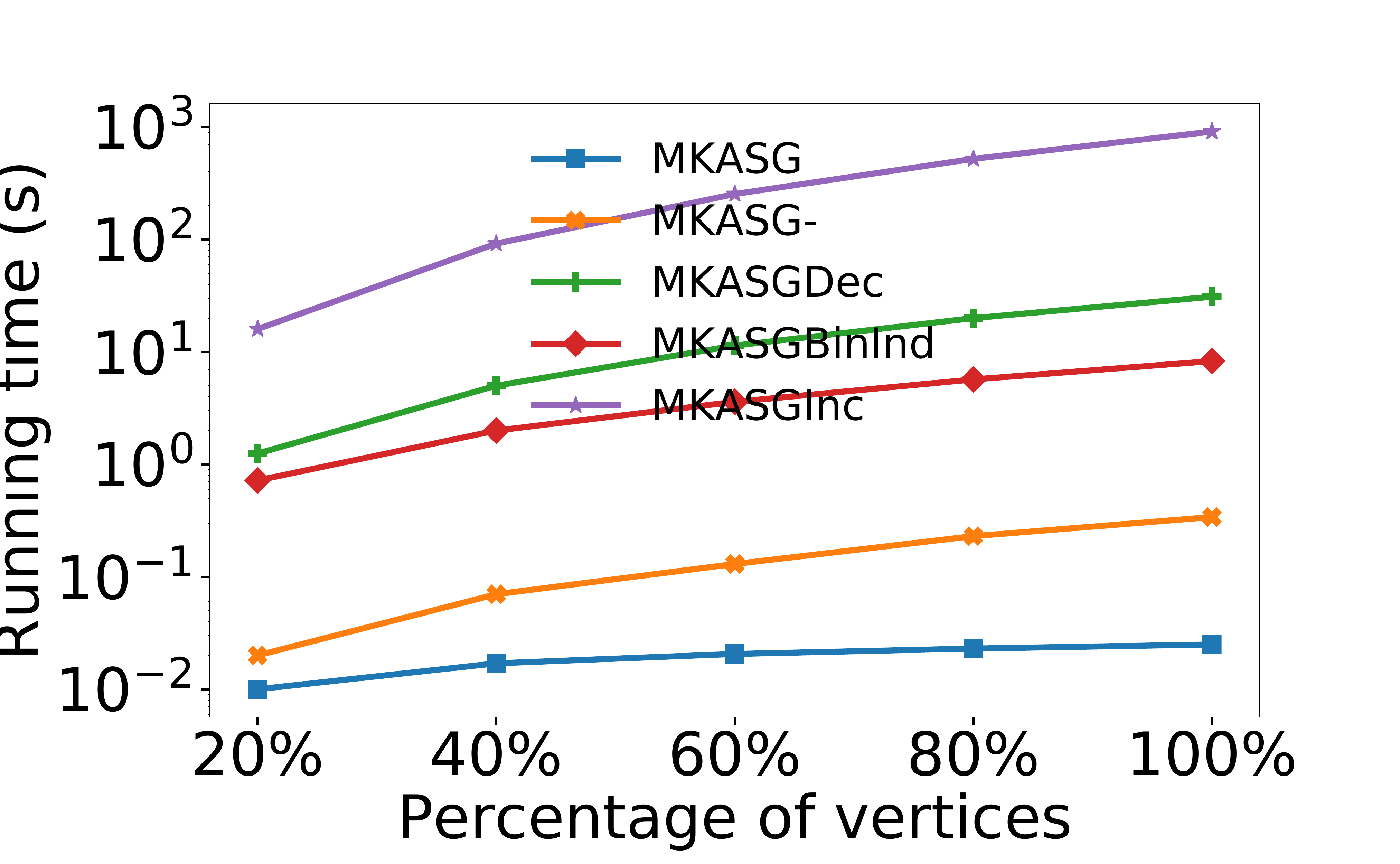}}
	\subfloat[Foursquare]{\includegraphics[width=3.7cm]{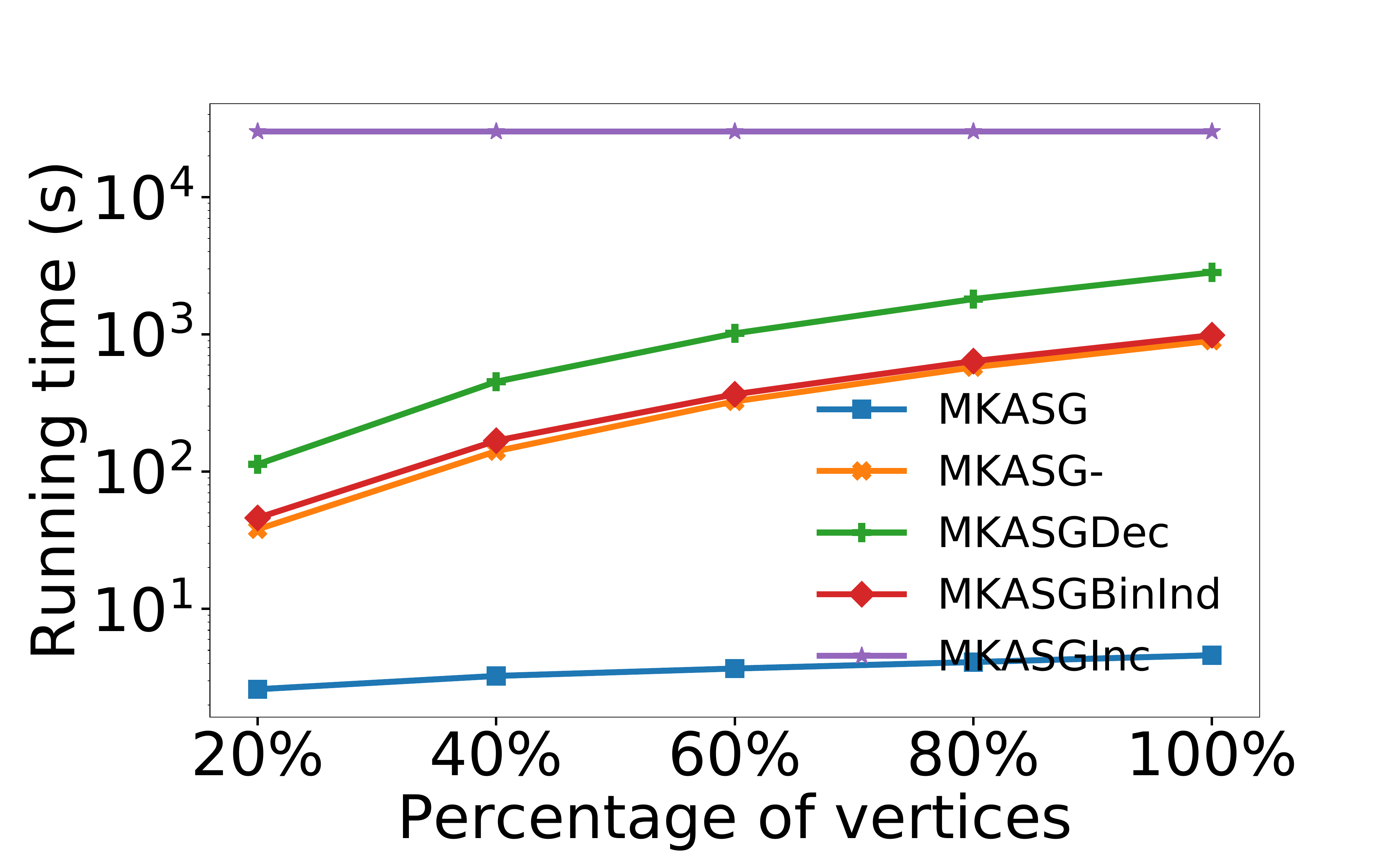}}
	\subfloat[Twitter]{\includegraphics[width=3.7cm]{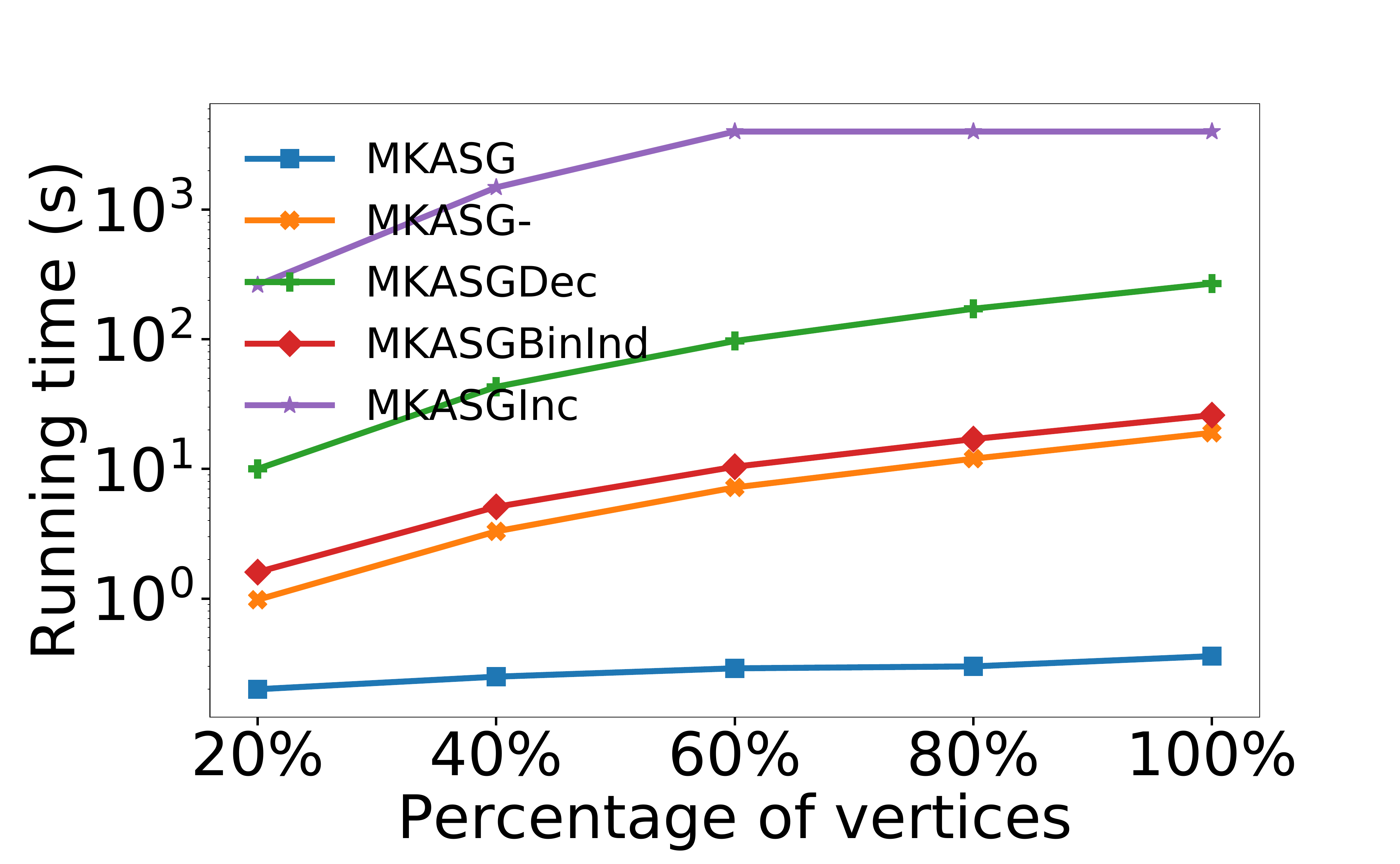}}
	\subfloat[Weibo]{\includegraphics[width=3.7cm]{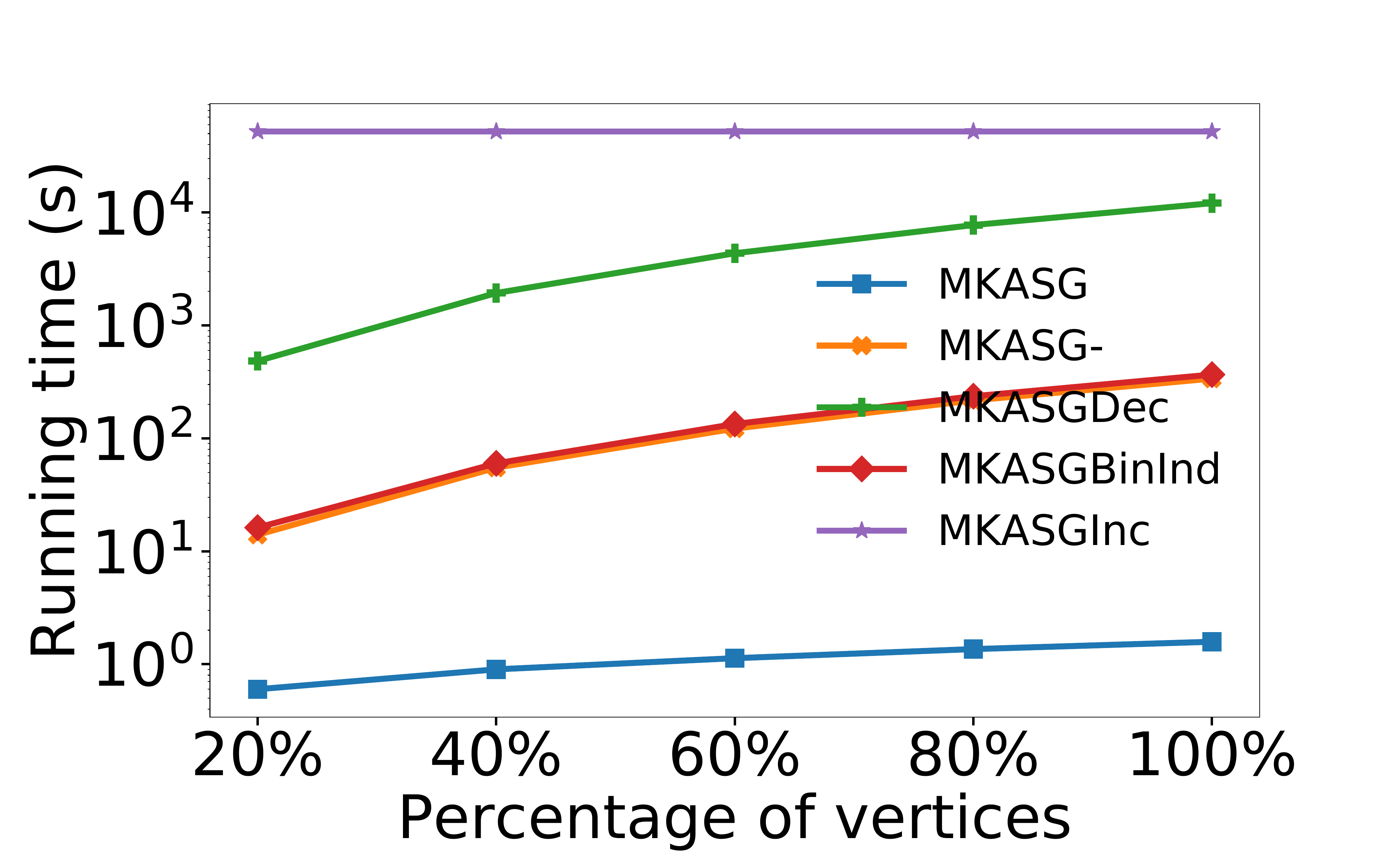}}

    \vspace{-11pt}

	\subfloat[Gowalla]{\includegraphics[width=3.7cm]{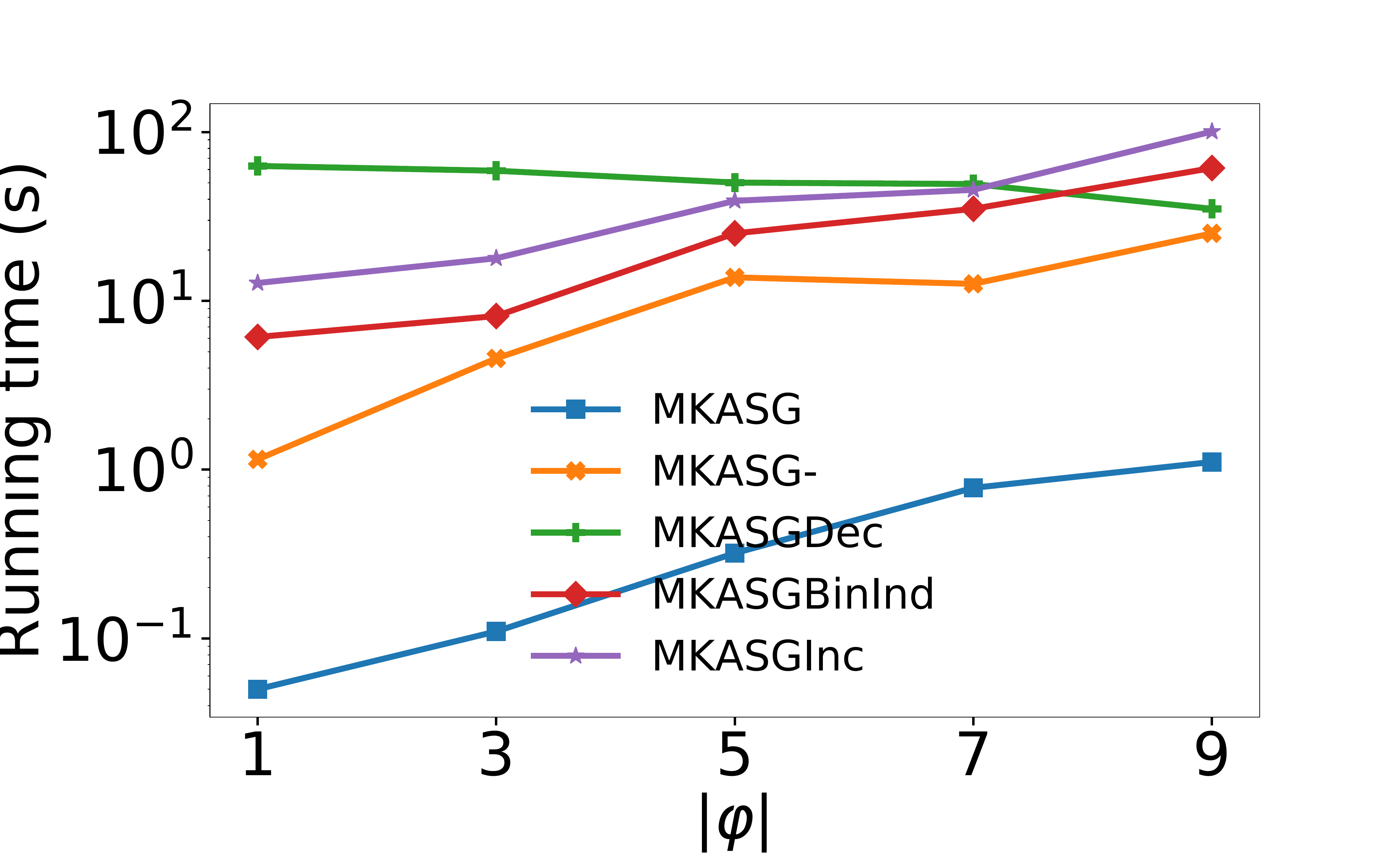}}
	\subfloat[Brightkite]{\includegraphics[width=3.7cm]{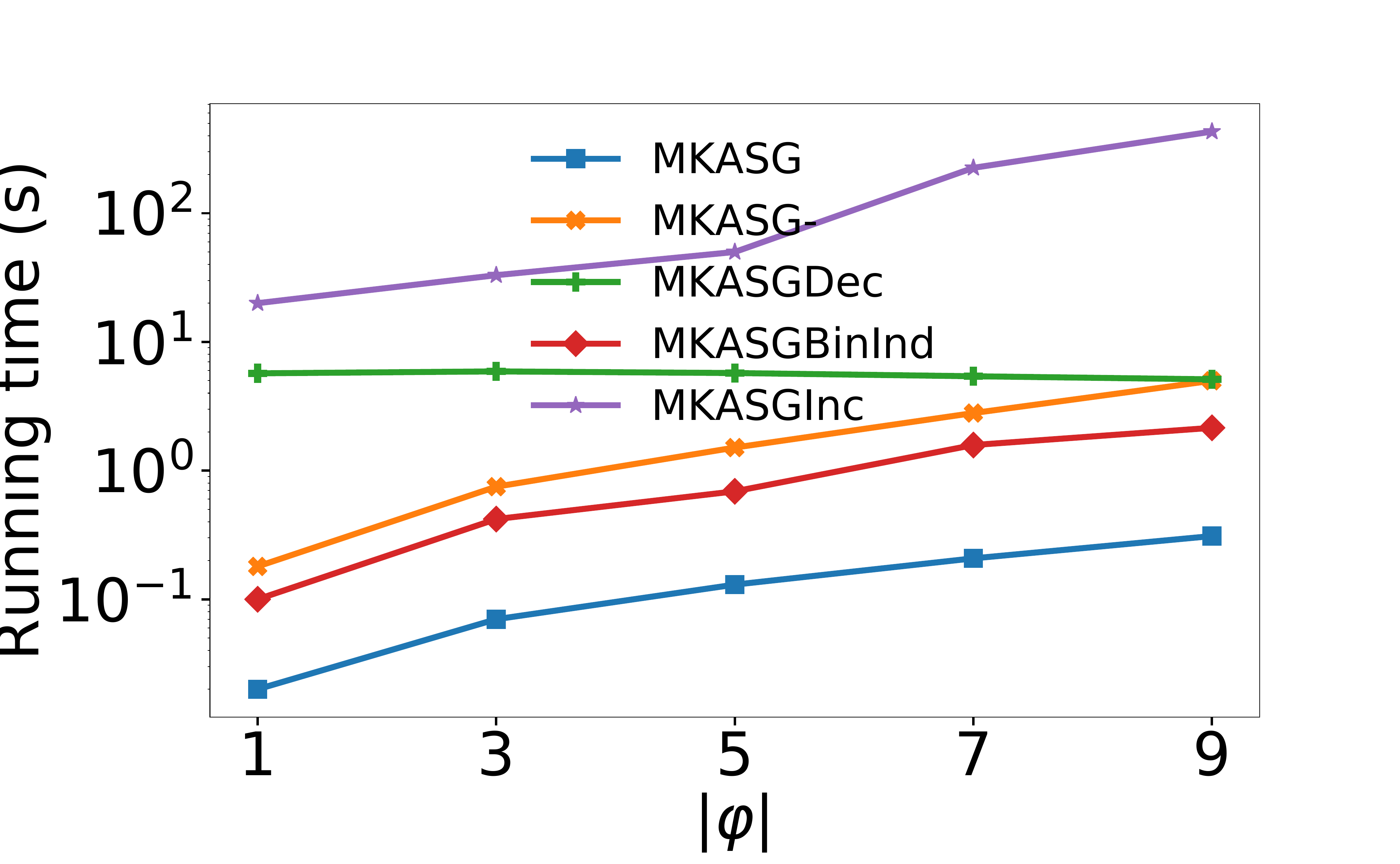}}
	\subfloat[Foursquare]{\includegraphics[width=3.7cm]{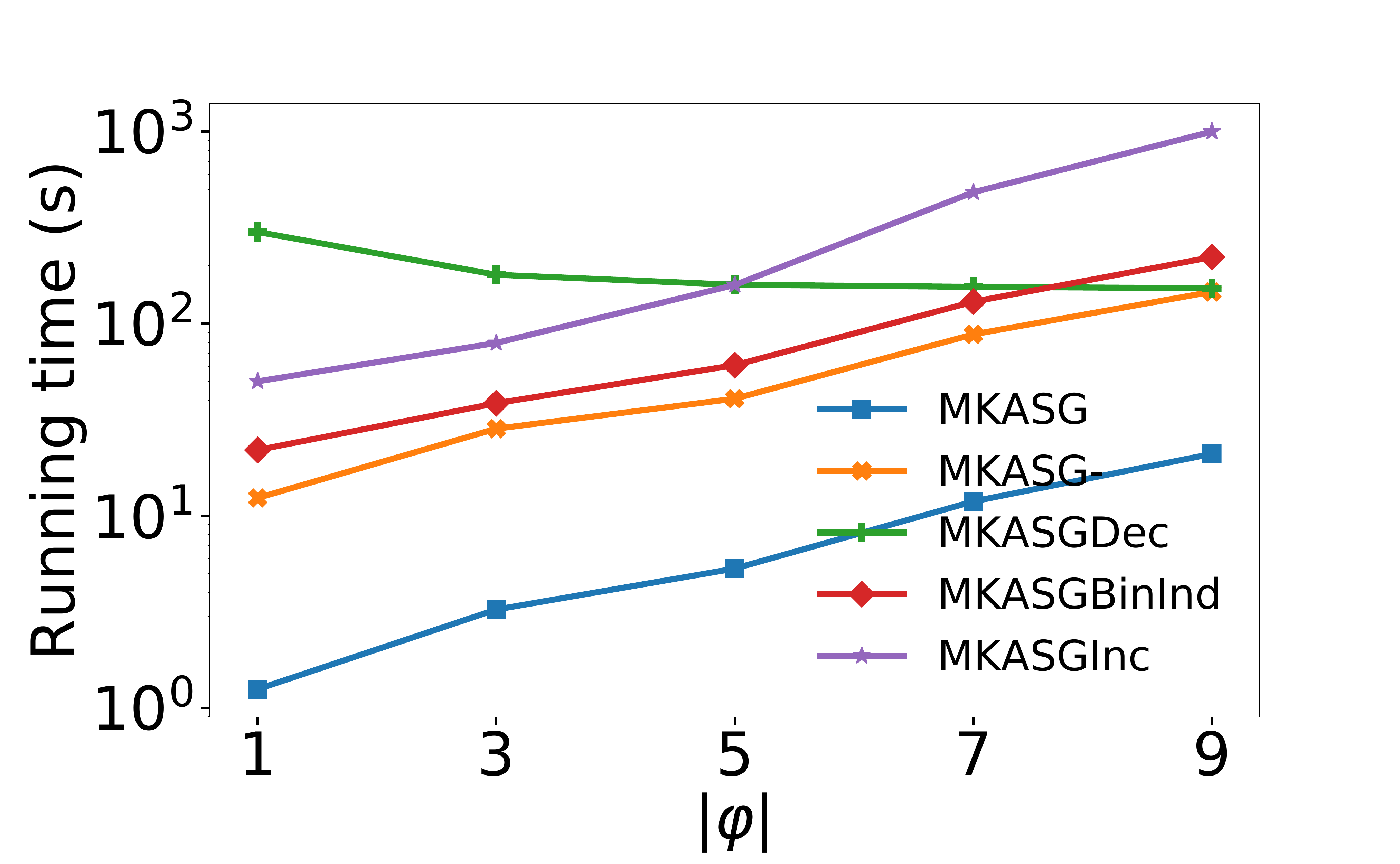}}
	\subfloat[Twitter]{\includegraphics[width=3.7cm]{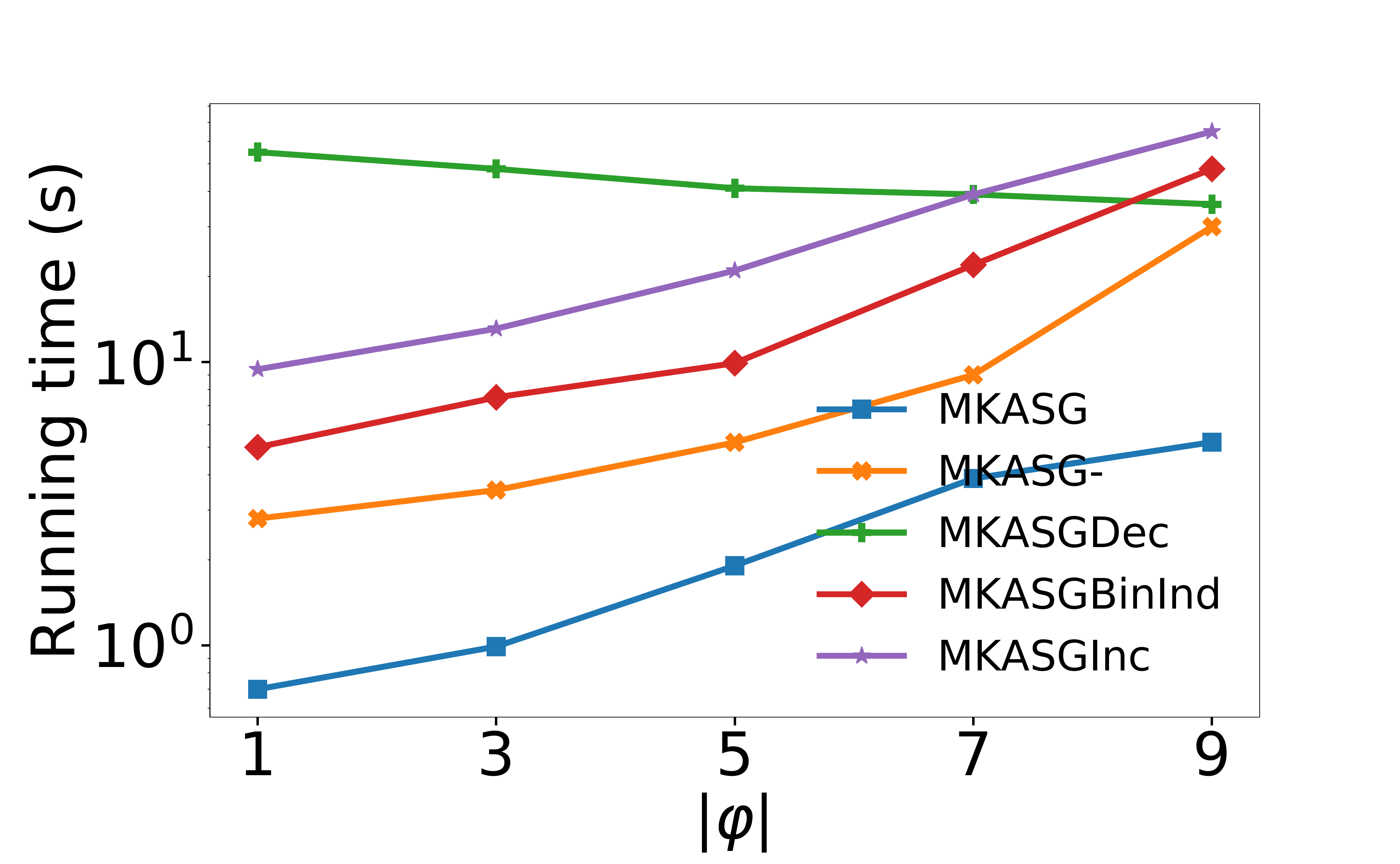}}
	\subfloat[Weibo]{\includegraphics[width=3.7cm]{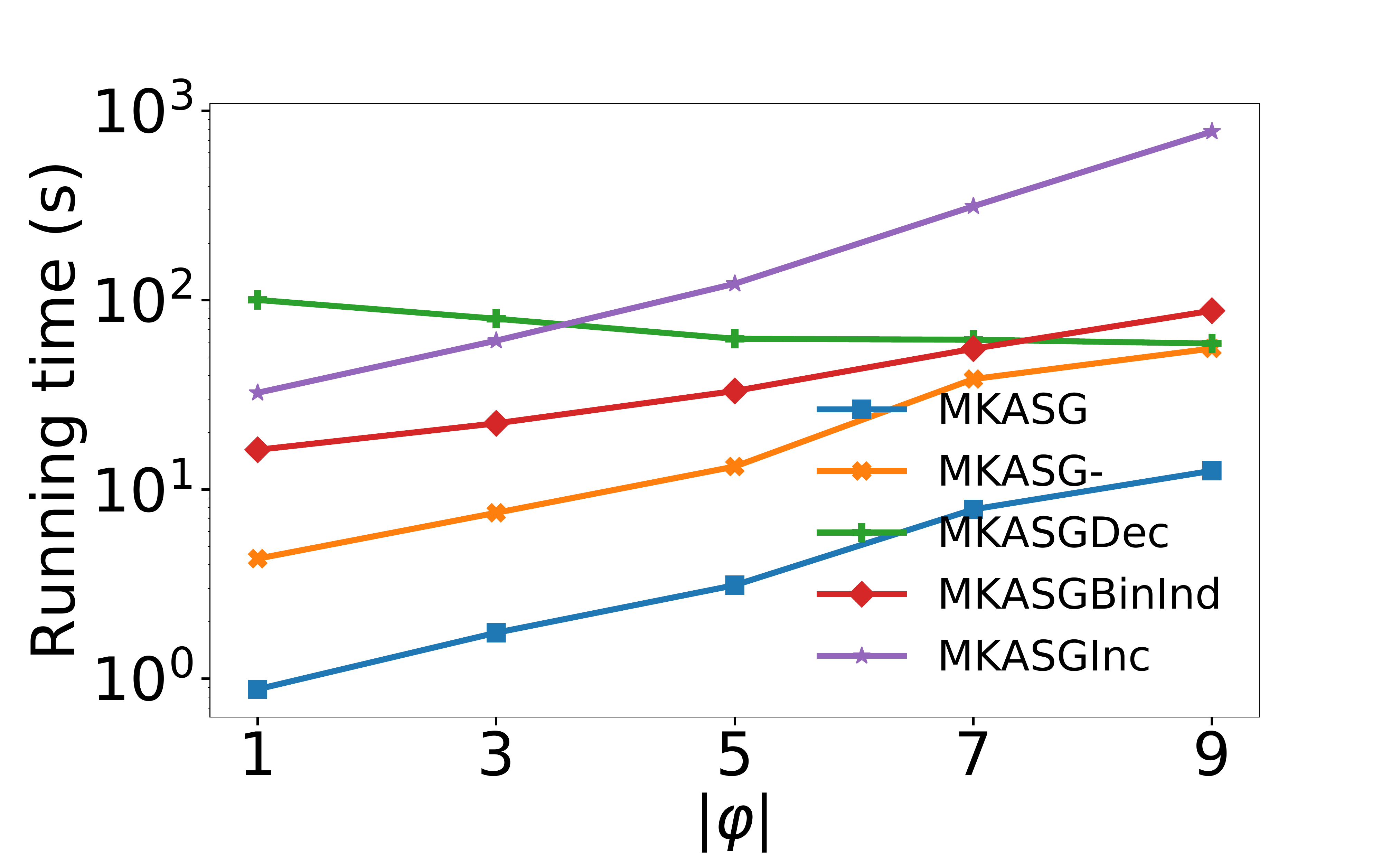}}

    \vspace{-11pt}

    \subfloat[Gowalla]{\includegraphics[width=3.7cm]{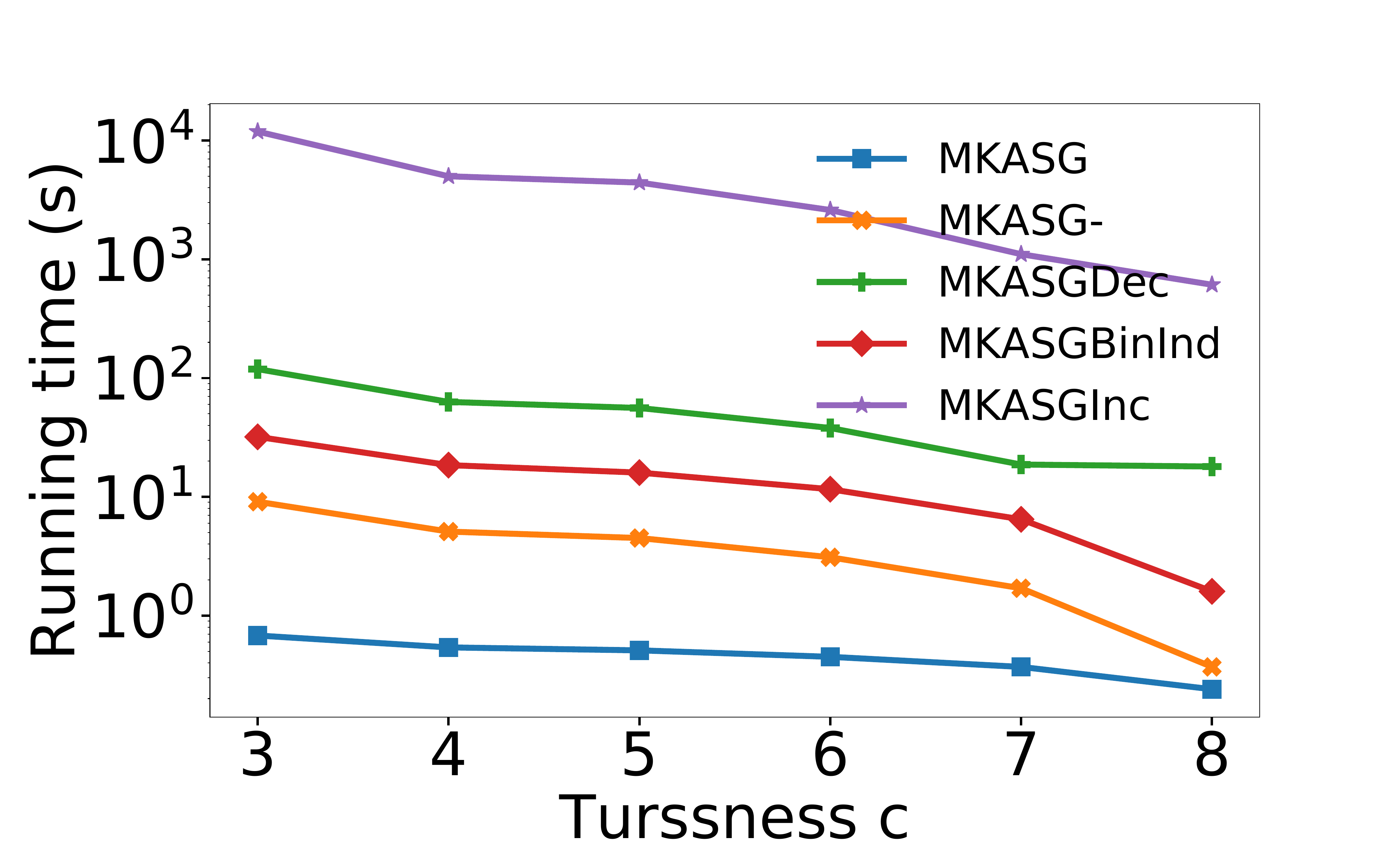}}
	\subfloat[Brightkite]{\includegraphics[width=3.7cm]{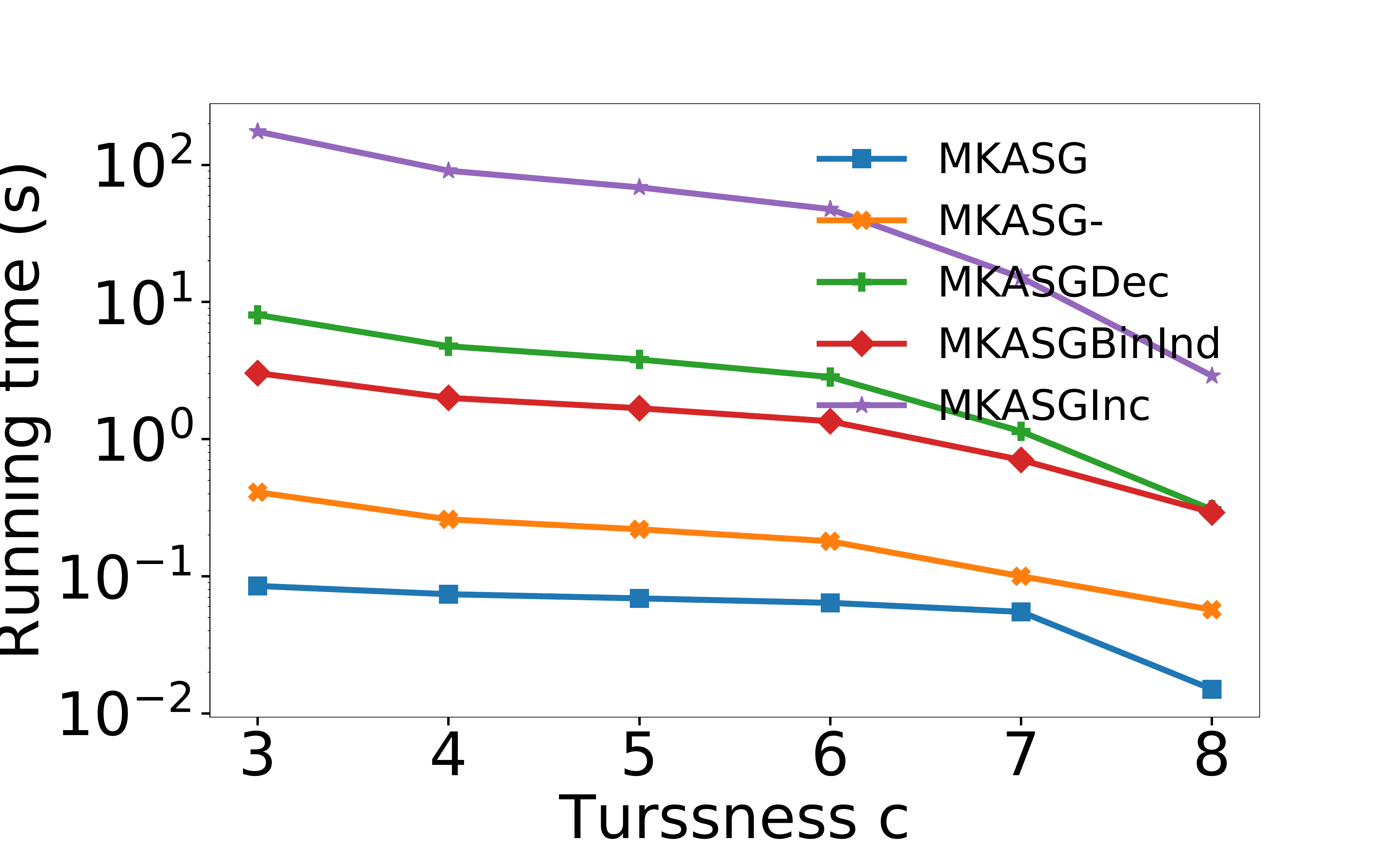}}
	\subfloat[Foursquare]{\includegraphics[width=3.7cm]{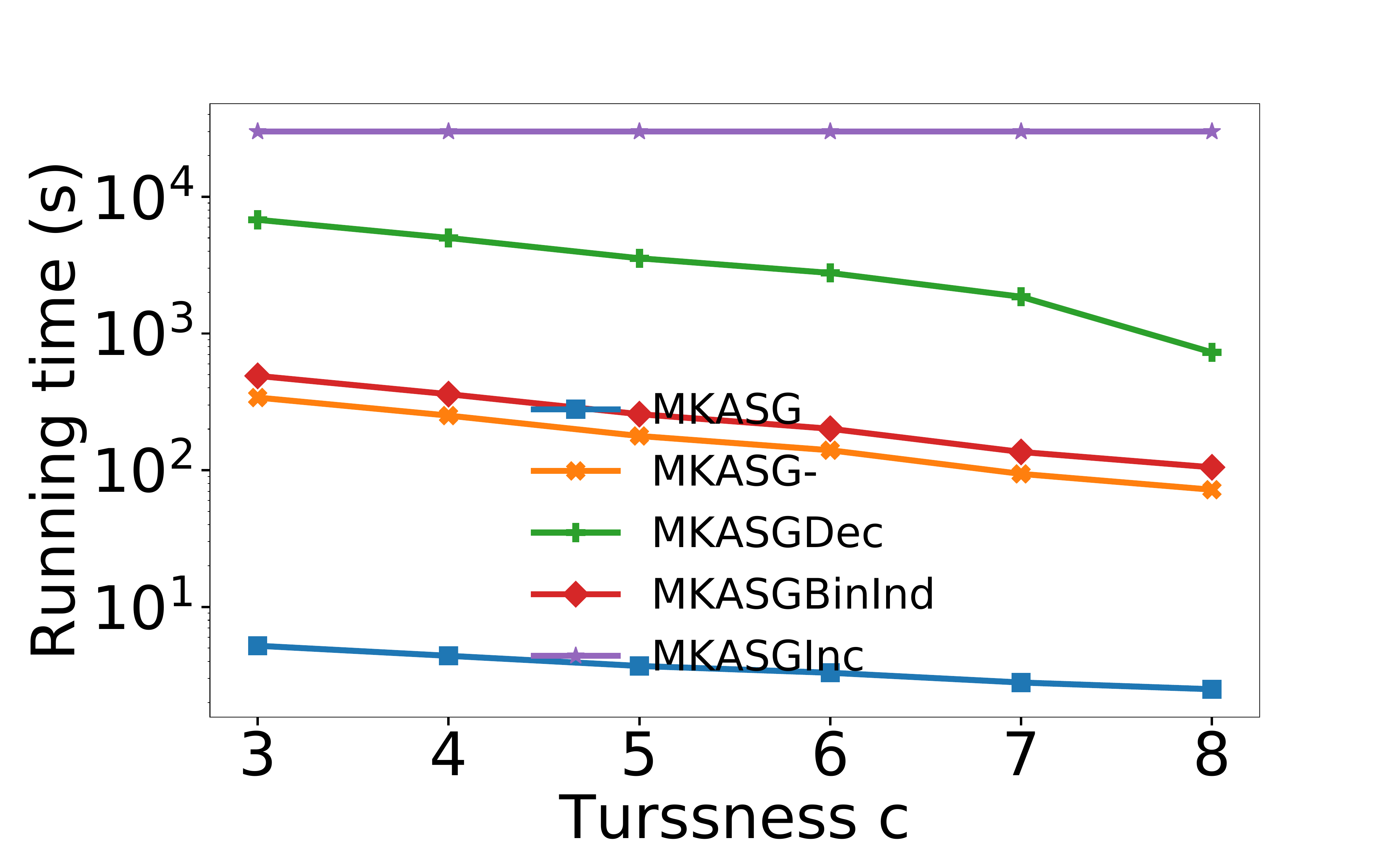}}
	\subfloat[Twitter]{\includegraphics[width=3.7cm]{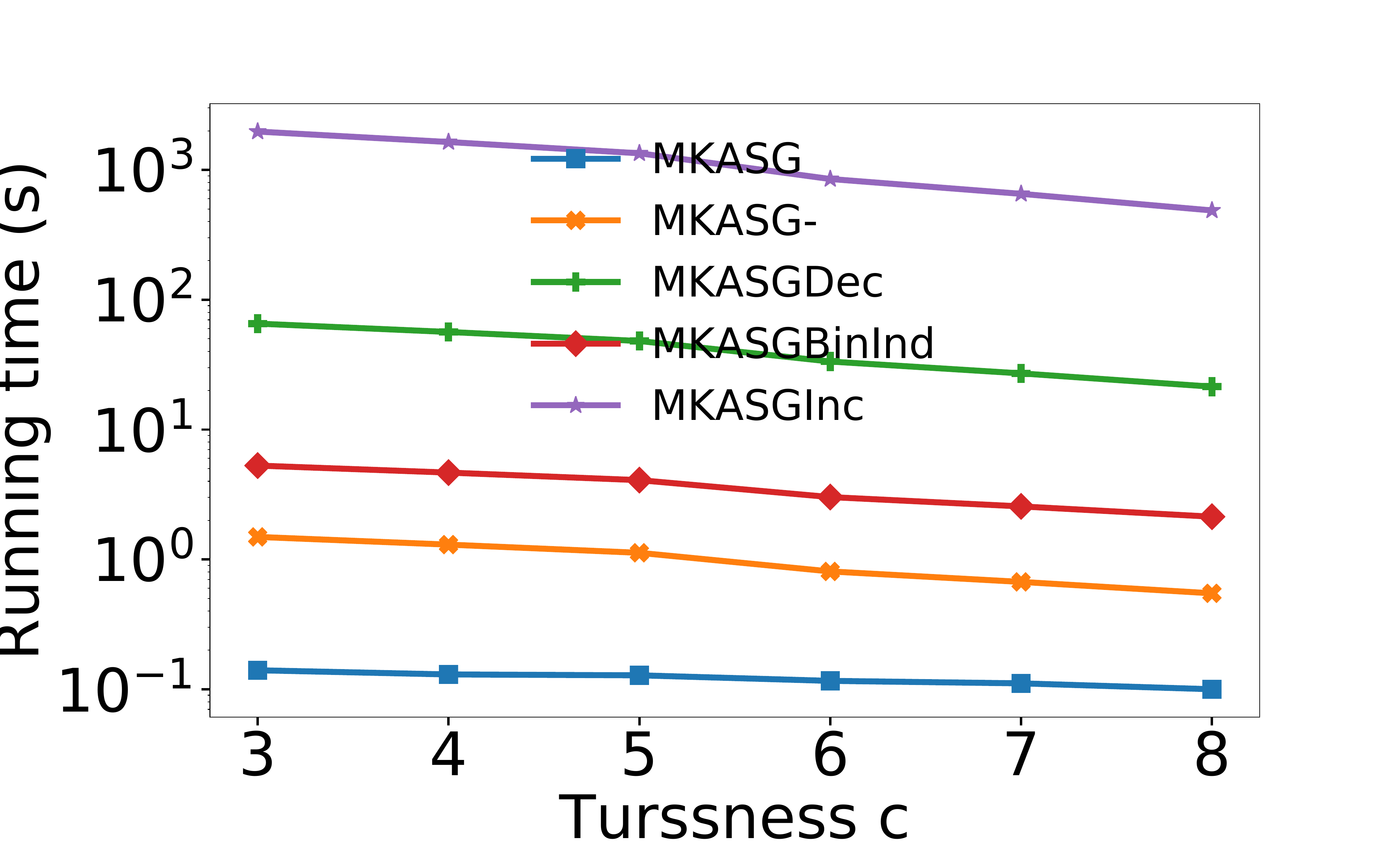}}
	\subfloat[Weibo]{\includegraphics[width=3.7cm]{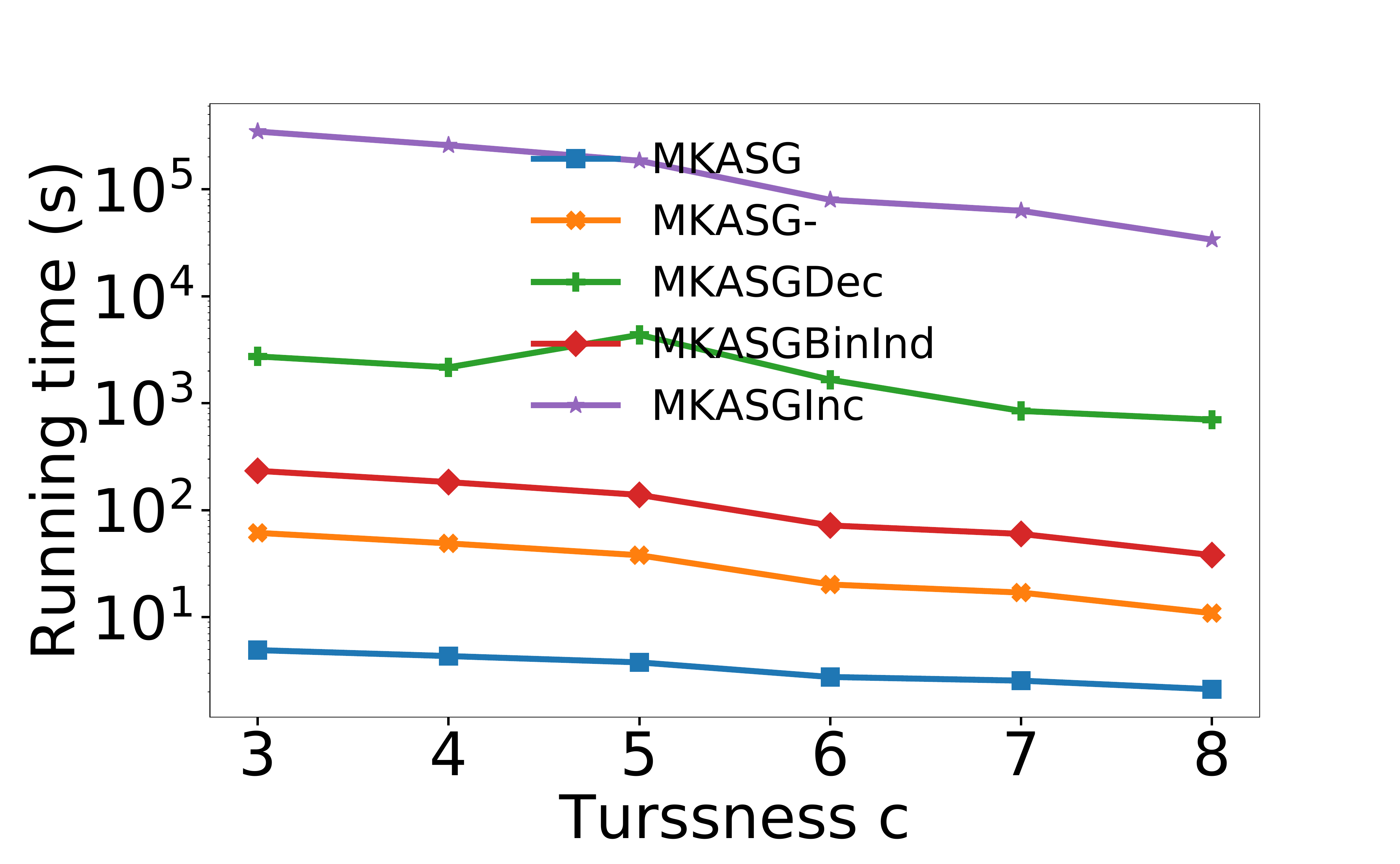}}

    \vspace{-11pt}

    \subfloat[Gowalla]{\includegraphics[width=3.7cm]{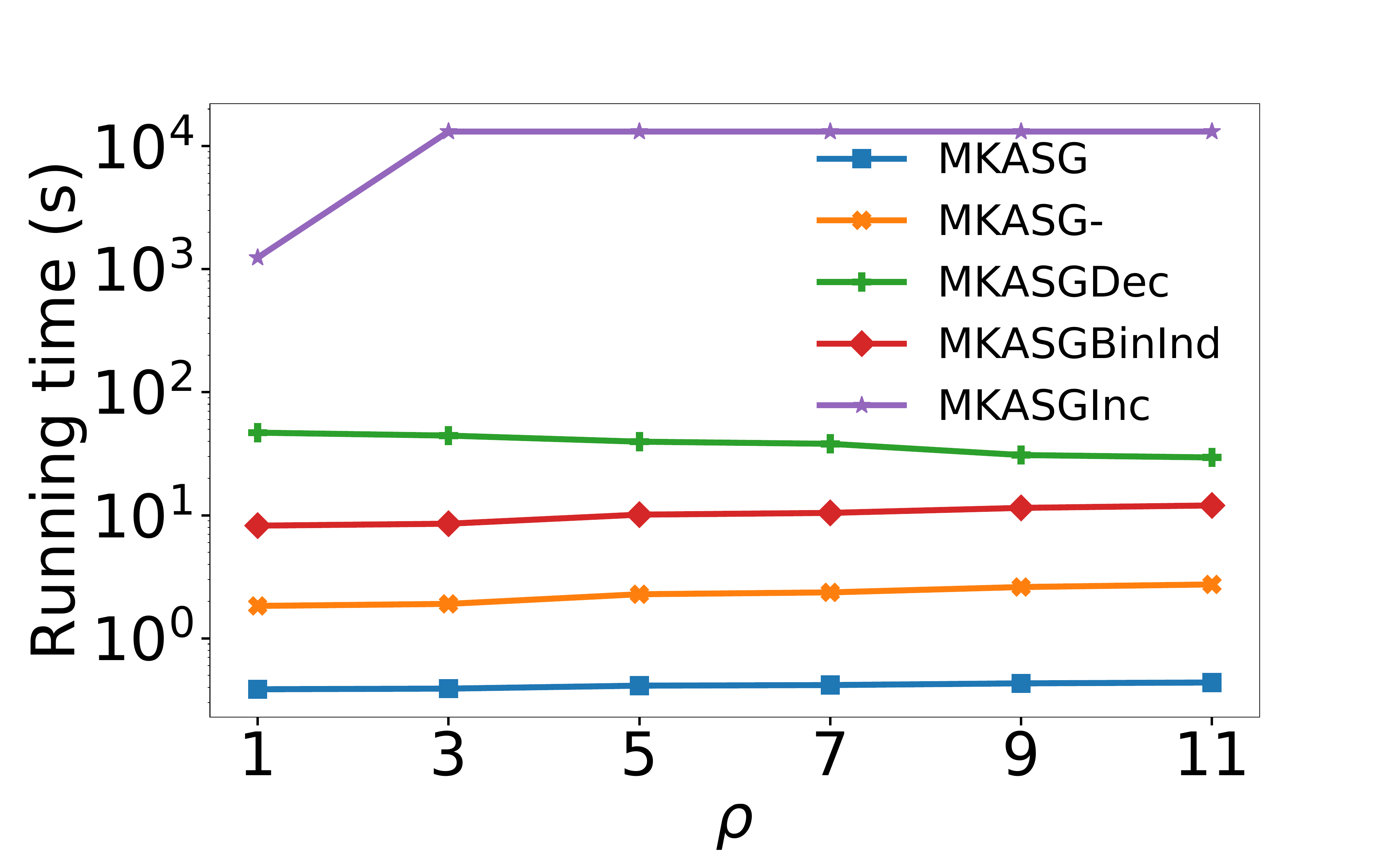}}
	\subfloat[Brightkite]{\includegraphics[width=3.7cm]{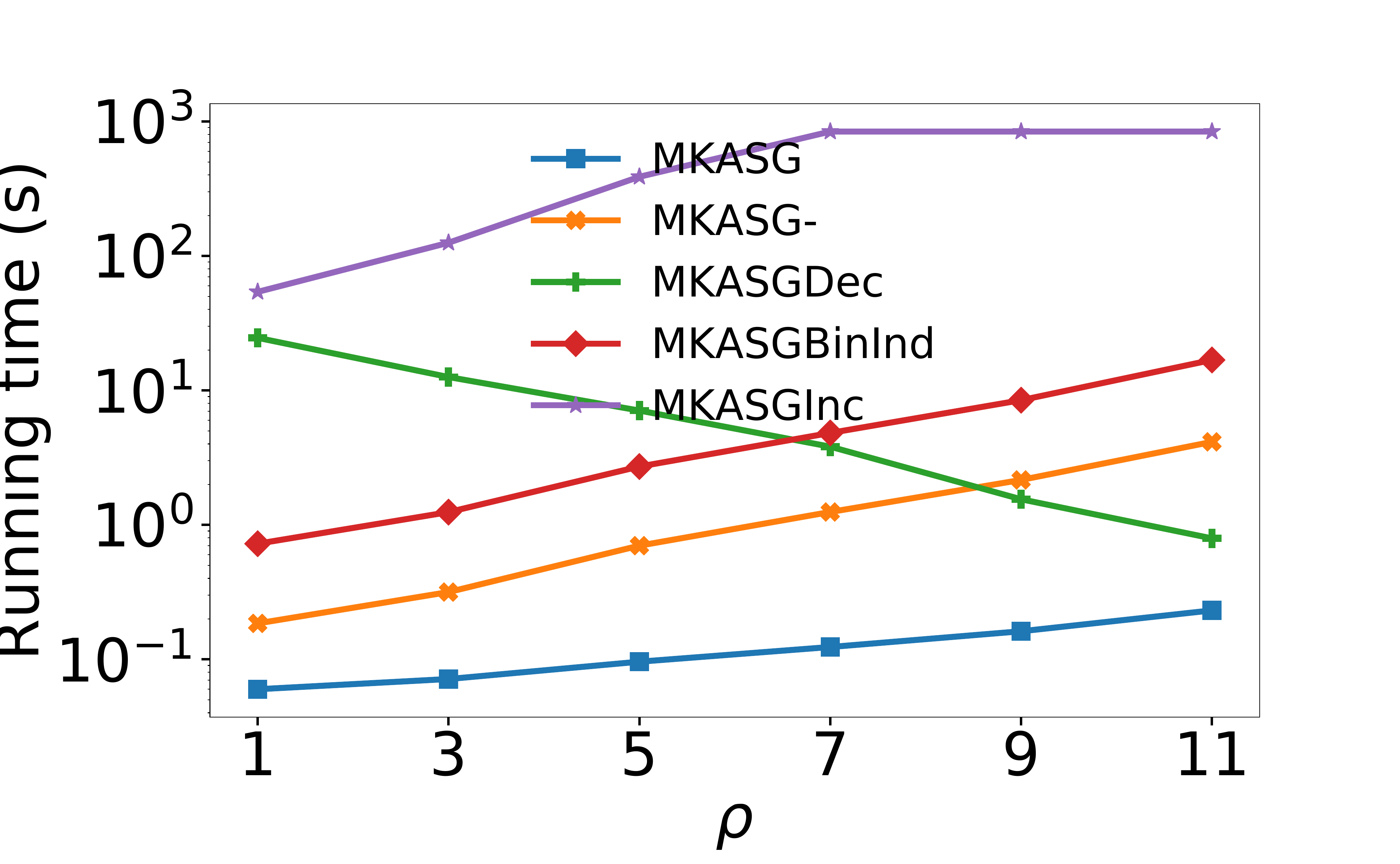}}
	\subfloat[Foursquare]{\includegraphics[width=3.7cm]{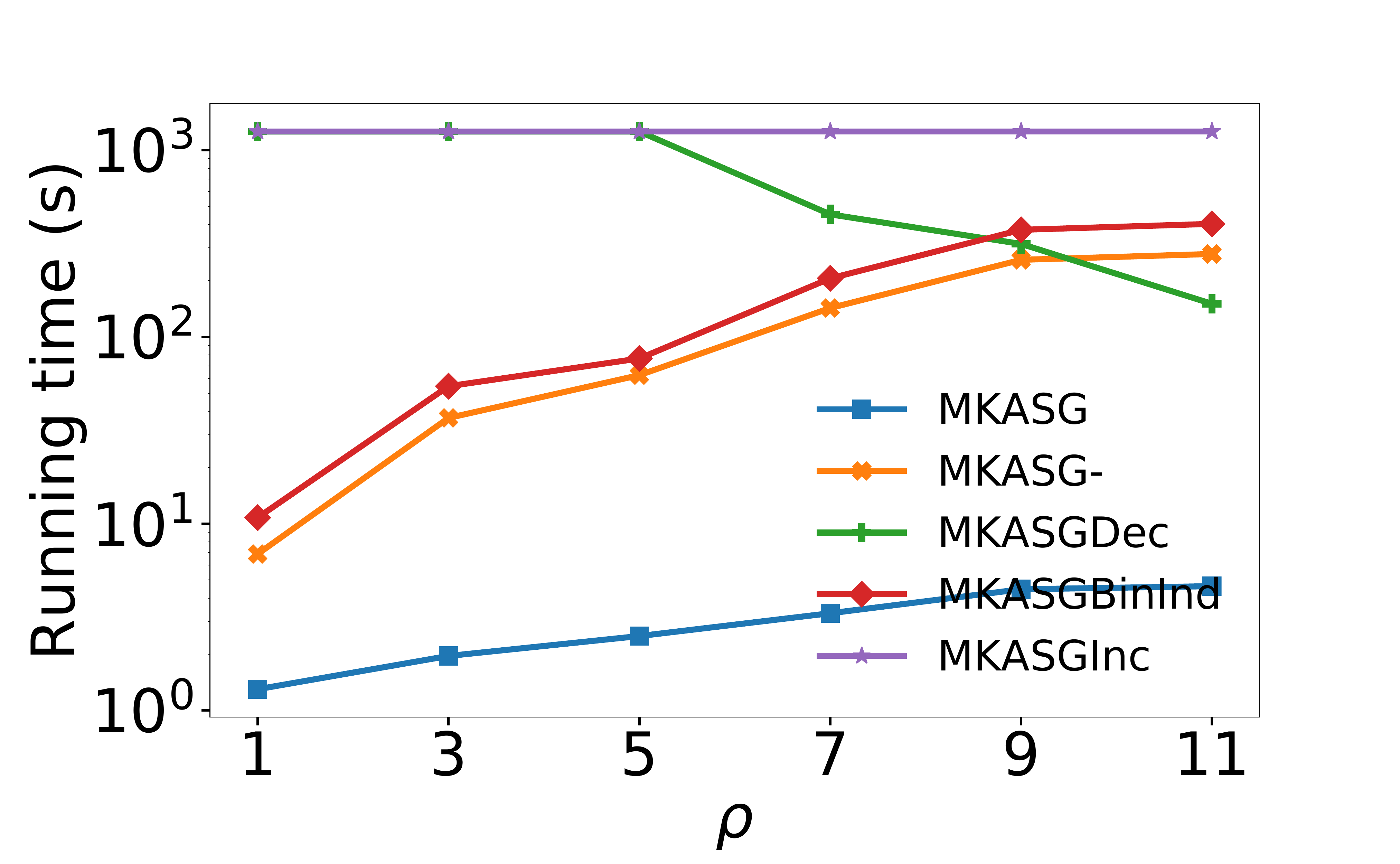}}
    \subfloat[Twitter]{\includegraphics[width=3.7cm]{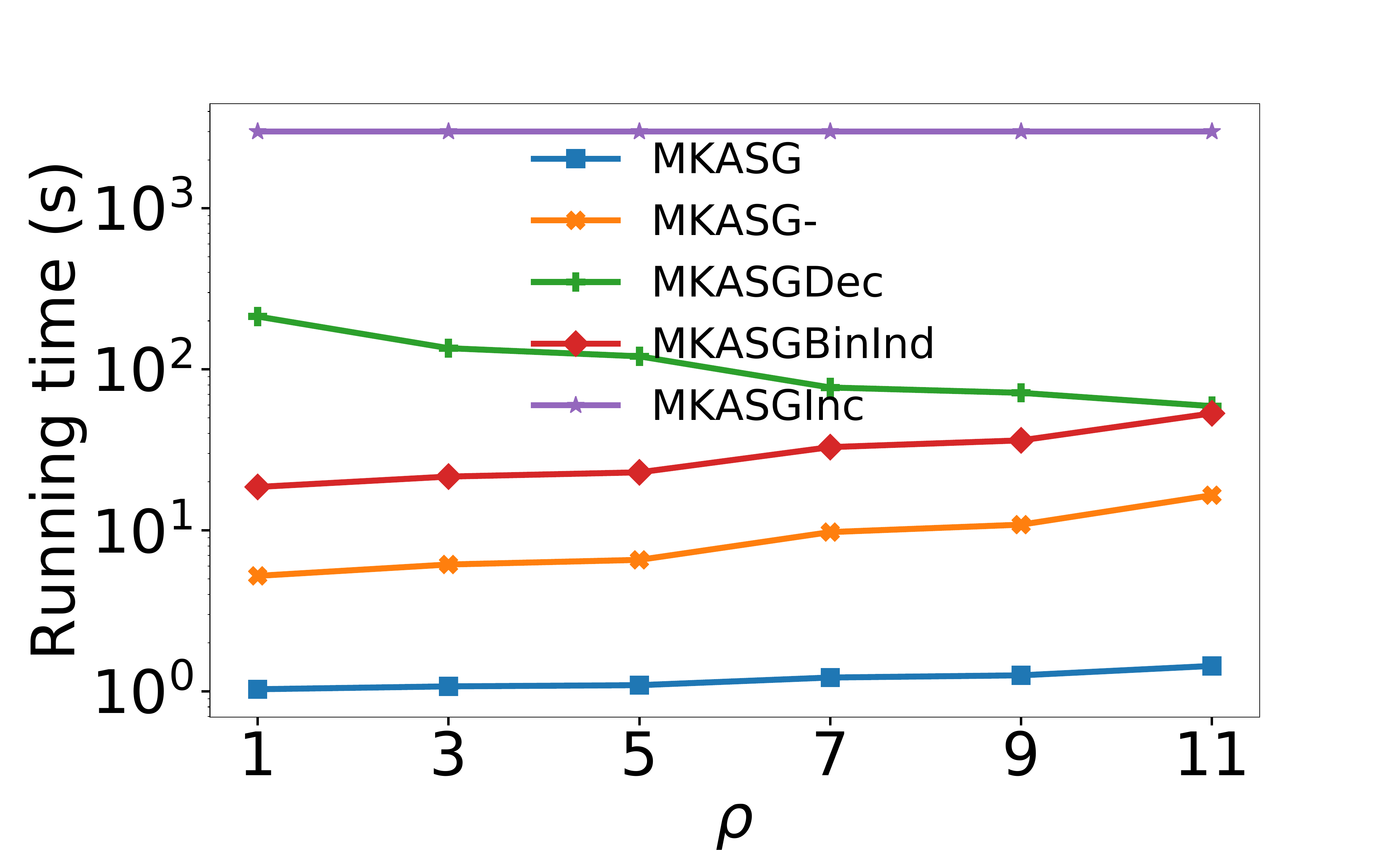}}
	\subfloat[Weibo]{\includegraphics[width=3.7cm]{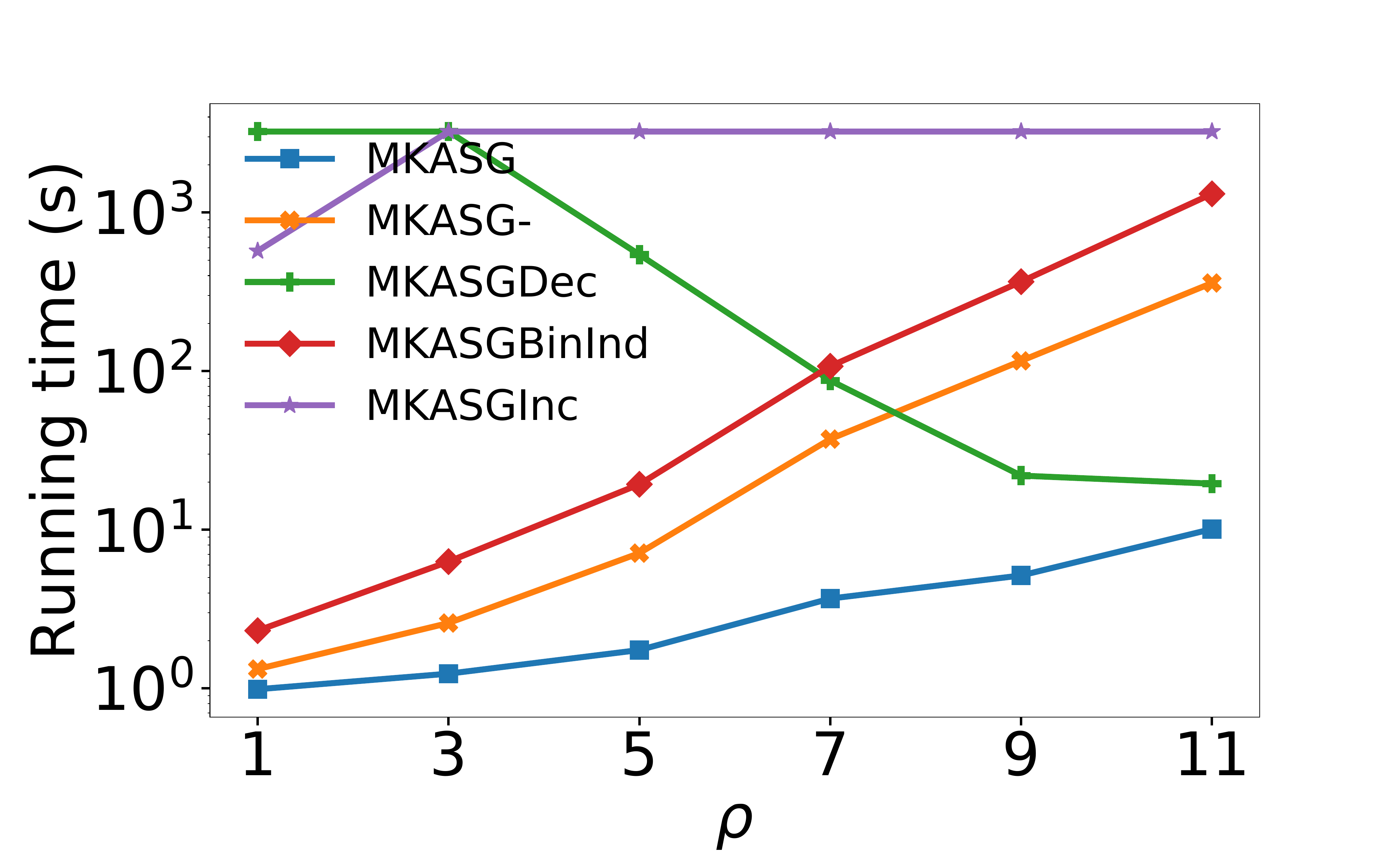}}

    \captionsetup{justification=centering}
	\caption{Efficiency evaluation}\label{fig:expeffi}
    \vspace{-5pt}
\end{figure*}

\vspace{-10pt}
\section{Experimental Study}\label{sec:exp}
In this section, we conduct experimental studies on real datasets to evaluate the proposed model and algorithms. We first discuss non-trivial baseline algorithms used in the experimental study.

\subsection{Evaluated Algorithms}
In the experiment we denote Algorithm~\ref{alg:eralg} as \textsf{MKASG}, the incremental approach as \algbaselineD, the decremental approach as \algbaselineB~and the binary search based approach with R-tree index as \algbaselineC.
Besides, we also consider a simplified Algorithm~\ref{alg:eralg} as one of the baselines discussed below.

\noindent\textbf{\algbaselineA}. This algorithm is a simplified version of \textsf{MKASG}. For the expanding stage, it only applies power law expansion and for the reducing stage it does not applies the proposed online index.
This is used to show the power of the search framework used in this paper.

For all the baseline approaches, we apply $(\rho,c)$-truss based prunings in prior.




\subsection{Experiment Setups}

\noindent\textbf{Datasets}. For efficiency evaluation, we conducted the experiments over five real social network datasets including Gowalla, Brightkite, Foursquare, Weibo and Yelp.
Each social user contains some check-in locations.
Table~\ref{tab:dataset} presents the statistics for all datasets.
Since we only need one check-in for each user, we select the latest check-in as the spatial coordinate for a vertex, if the user has multiple check-ins.
The keyword attribute of each user is randomly assigned for the first four datasets, which refers to the current main interest.
The Yelp contains real social relationships, check-ins and textual information.
For effectiveness evaluation, we use WoW dataset. WoW is player data in World of Warcraft game. The social network in WoW is friendships over players in game, the spatial information is location in the virtual game world, and the keywords are players' real classes and roles in the game.

\noindent\textbf{Parameter settings}.
The experiments are evaluated using different settings of query parameters: $c$ (the minimum truss number), reasonable sets of keywords $\qk$ as well the keyword constraint parameter $\rho$. The query locations are generated randomly.
The ranges of the parameters and their default values are shown in Table~\ref{tab:para}, in which we select reasonable $c$ based on datasets.
Furthermore, when we vary the value of a parameter for evaluation, all the other parameters are set as their default values.

All algorithms are implemented in C++, and the experiments are conducted on a PC with CPU of AMD 3900x (12 cores, 24 threads), memory of 128GB DDR4 3600HZ, and Windows 10 (build 1803). All experiments are conducted no less than 100 times and the average results are demonstrated.


\subsection{Efficiency Evaluation}

\noindent\textbf{Scalability}. To verify the scalability of our algorithms, we choose different sizes of sub-datasets by selecting different percentages of vertices in each dataset.
The results are displayed in Figures~\ref{fig:expeffi}(a) and (e). Overall speaking, algorithms using our proposed search framework (\algbaselineA, \textsf{MKASG}) are more scalable compared to \algbaselineB, \algbaselineD, and \algbaselineC.
This is because the proposed search framework has nice property that can limit the search region while preserving optimum result. \textsf{MKAS} is the one most scalable since it incorporates with the proposed techniques which make the time complexity of \textsf{MKASG} optimal.
On the other hand, \algbaselineD~is the least scalable due to its high time complexity.
In large dataset Foursquare, it cannot get result over 24 hours.
For \algbaselineC~using R-tree, it is slower than the algorithms based on our proposed search framework. It seems to counterintuitive since using index reduces the search space explored.
This is because the R-tree based index can only locate vertex efficiently; however, to identify the subgraph and trussness of the subgraph in a region, it has to perform induced subgraph and truss computations repeatedly and the total repeated computations are worse than the search framework proposed in this paper.

\noindent\textbf{Varying $|\qk|$}. Figures~\ref{fig:expeffi}(f) to (j) demonstrate the running times as $|\qk|$ varies for different datasets.
As the number of query keywords increases, the running time for \textsf{MKASG}, \algbaselineA, \algbaselineC~and \algbaselineD~rises.
This is because having more keywords indicates more data need to be explored by those algorithms since they explore vertices from the region near by the query location to the region containing the optimum result.
For \algbaselineB, more keywords lead to less computations since less vertices need to be deleted, which makes its running time decrease for all datasets as $|\qk|$ increases.
\textsf{MKASG} outperforms all the other algorithms substantially.
\textsf{MKASG} can find optimum result within $1$ second in most of the datasets while can still answer a query for extreme large dataset in reasonable time, i.e., in a few seconds.
For most of the dataset, \textsf{MKASG-} has the second best performance given the evaluated parameters. This shows the power of the proposed search framework. This is because \textsf{MKASG-} can bound repeated computation nicely.

\noindent\textbf{Varying $c$}. We evaluate the performance for all the algorithms when varying the trussness $c$ in Figures~\ref{fig:expeffi}(k) and (o).
In general, as $c$ increases, the running time for all the algorithms reduces.
The reason is that the size of subgraph with high $c$ tends to be small, which makes search space decrease as $c$ rises.
Noticeablely, \textsf{MKASG} outperforms other algorithms in most of the datasets. Especially for Brightkite and Twitter, it can get result in less than $1$ second.
Again, this experiment also justifies the superiority of the proposed search framework, i.e., \textsf{MKASG-} is the second faster for all datasets.
\textsf{MKASGInc} is slower than all the other algorithms and it runs over 24 hours for Foursquare to get the result.
This is because its highly repeated computations, which makes it worse than \textsf{MKASGBinInd} and \textsf{MKASGDec}.
At last, compared to other algorithms, \textsf{MKASG} is less sensitive to the changes of $c$ because of its optimality.

\noindent\textbf{Varying $\rho$}. In Figures~\ref{fig:expeffi}(p) to (t), the running time for the algorithms when we change $\rho$ are shown for different datasets.
For all datasets, the running time of all algorithms increases as $\rho$ increases except for \textsf{MKASGInc}. The reason is similar to what has been explained when varying $|\qk|$.
For both Gowalla and Weibo, \textsf{MKASG} is not very sensitive to the change of $\rho$.
This is because for all these datasets the proposed initial search bound can approach to optimum result effectively, and the dominating computation is just trussness verification.
This set of experiments also demonstrate the power of the proposed search framework, i.e., both \textsf{MKASG} and \textsf{MKASG-} outperform other algorithms clearly for Gowalla, Brightkite, Twitter and Weibo.
Compared to \textsf{MKASGBinInd}, \textsf{MKASGDec} is much slower. This is because two reasons: first \textsf{MKASGDec} takes extra cost for sorting and secondly \textsf{MKASGDec} does not have social prunings.

\noindent\textbf{Pruning effectiveness evaluation}. We show pruning effectiveness in Table~\ref{tab:pruning} in term of size ratio for corresponding subgraphs evaluated by \textsf{MKASG}.
The result is the average of $200$ randomly generated queries with default settings but different query keywords for every dataset.
As we can see, the maximal $(\rho,c)$-truss based pruning can filter out 40\% to 60\% of vertices from the original graph.
Using the power law expanding,  our algorithm only evaluates 20\% to 35\% of maximal $(\rho, c)$-trusses for corresponding datasets.
It is very noticeable that, our proposed $\rho$ potential and truss potential subgraphs for different datasets are extremely small.
This further justifies the effectiveness of our proposed pruning techniques.

\begin{table}[t]
\scriptsize
\centering
	\begin{tabular}{|r|r|r|r|r|}
		\hline
		Dataset & $\frac{|H|}{|G|}$  &  $\frac{|\Delta H_{\le d^{*}}|}{|H|}$   & $\frac{|P_{\le d^{*}}|}{|\Delta H_{\le d^{*}}|}$  & $\frac{|C_{\le d^{*}}|}{|P_{\le d^{*}}|}$   \\
		\hline
		Gowalla & 58.4\% & 33.5\% & 12.4\% & 5.2\% \\
		\hline
		Brightkite & 47.6\% & 27.8\% & 17.42\% & 7.6\% \\
		\hline
		Foursquare & 39.5\% & 32.2\% &	10.3\% & 3.7\% \\
		\hline
		Weibo & 43.8\% & 21.2\% & 8.7\% & 4.3\% \\
		\hline
		Yelp & 57.2\% & 32.1\% & 11.3\% & 2.2\% \\
        \hline
		\end{tabular}
	\caption{Pruning evaluations}~\label{tab:pruning}
	\vspace{-5pt}
\end{table}

\subsection{Effectiveness Evaluation}
In this section, we report two case studies conducted to justify the effectiveness of the proposed model on WoW dataset.

\noindent\textbf{Data collection}. We collect friend list for players in a guild (similar to a community) in world of warcraft. Each player has two sets of attributes. The universes of the two sets are: $class:\{Warrior, Hunter, Rogue$, $\ldots,$ $Priest, Mage \}$ and $role:\{Damage, Healer, Tank\}$.

\noindent\textbf{Methodology}. In the virtual world, there are random missions requested in real time at a specific location.
There are two scenarios of popular missions. The first one needs a team containing 5 players and the second one need a team containing 15 players.
We compare the team formed by world of warcraft and team found by our algorithm.

\begin{figure}[t]
    \vspace{-8pt}
    \center
     \subfloat[WoW]{\includegraphics[width=3.2cm]{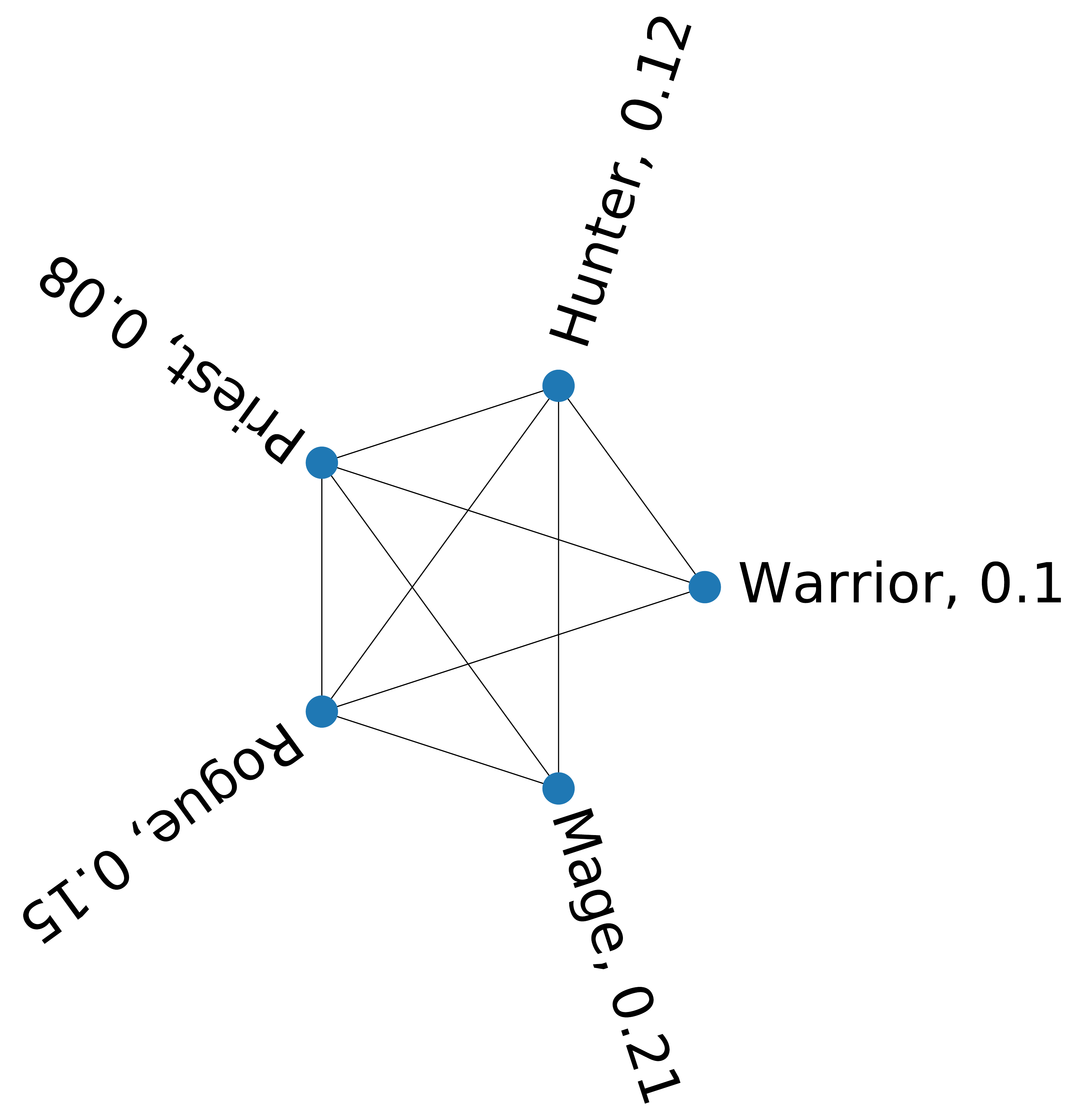}}
     \subfloat[WoW]{\includegraphics[width=3.2cm]{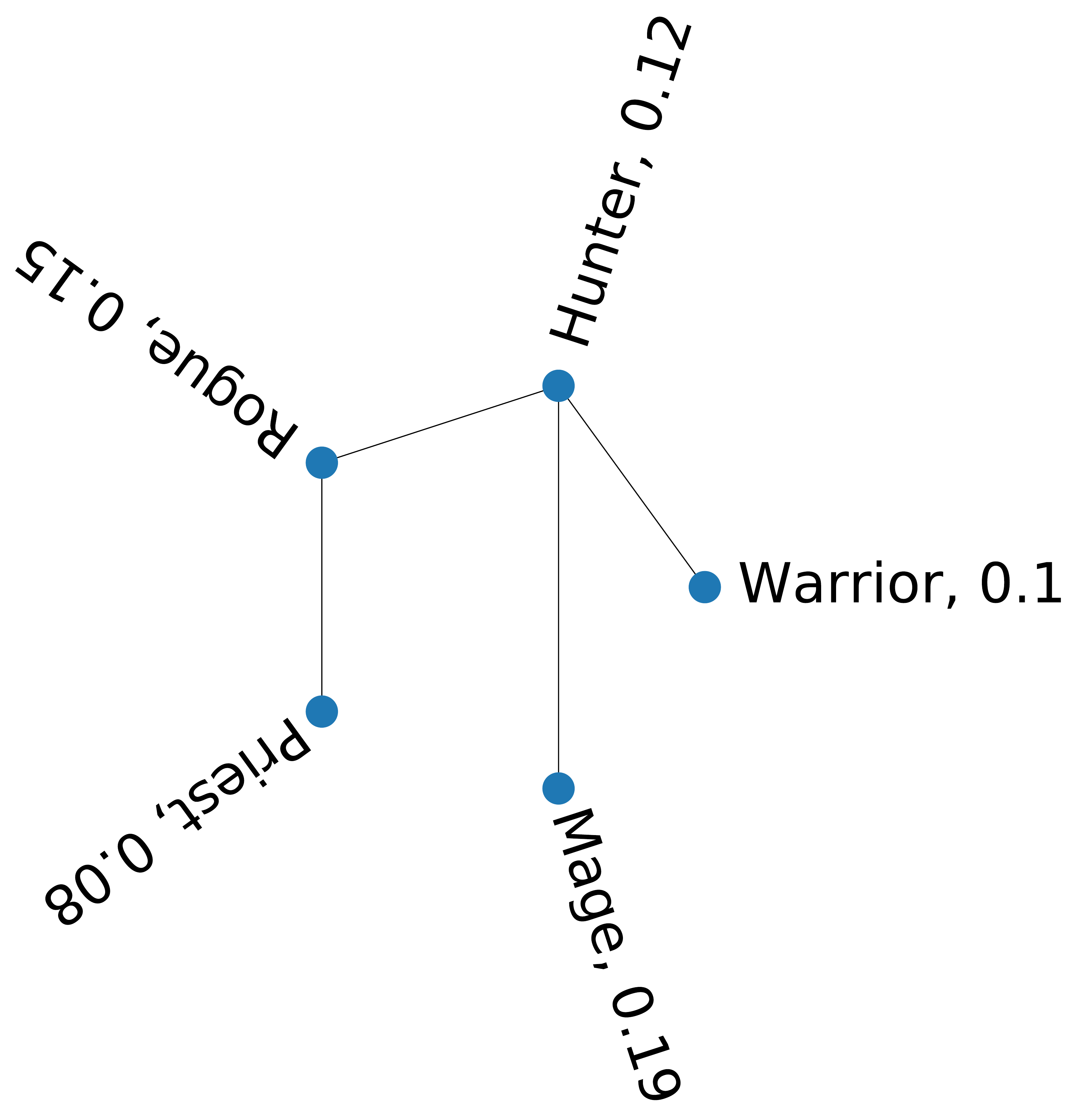}}
     \caption{Formulating small team}\label{fig:st}

        \vspace{5pt}
     \subfloat[WoW]{\includegraphics[width=3.2cm]{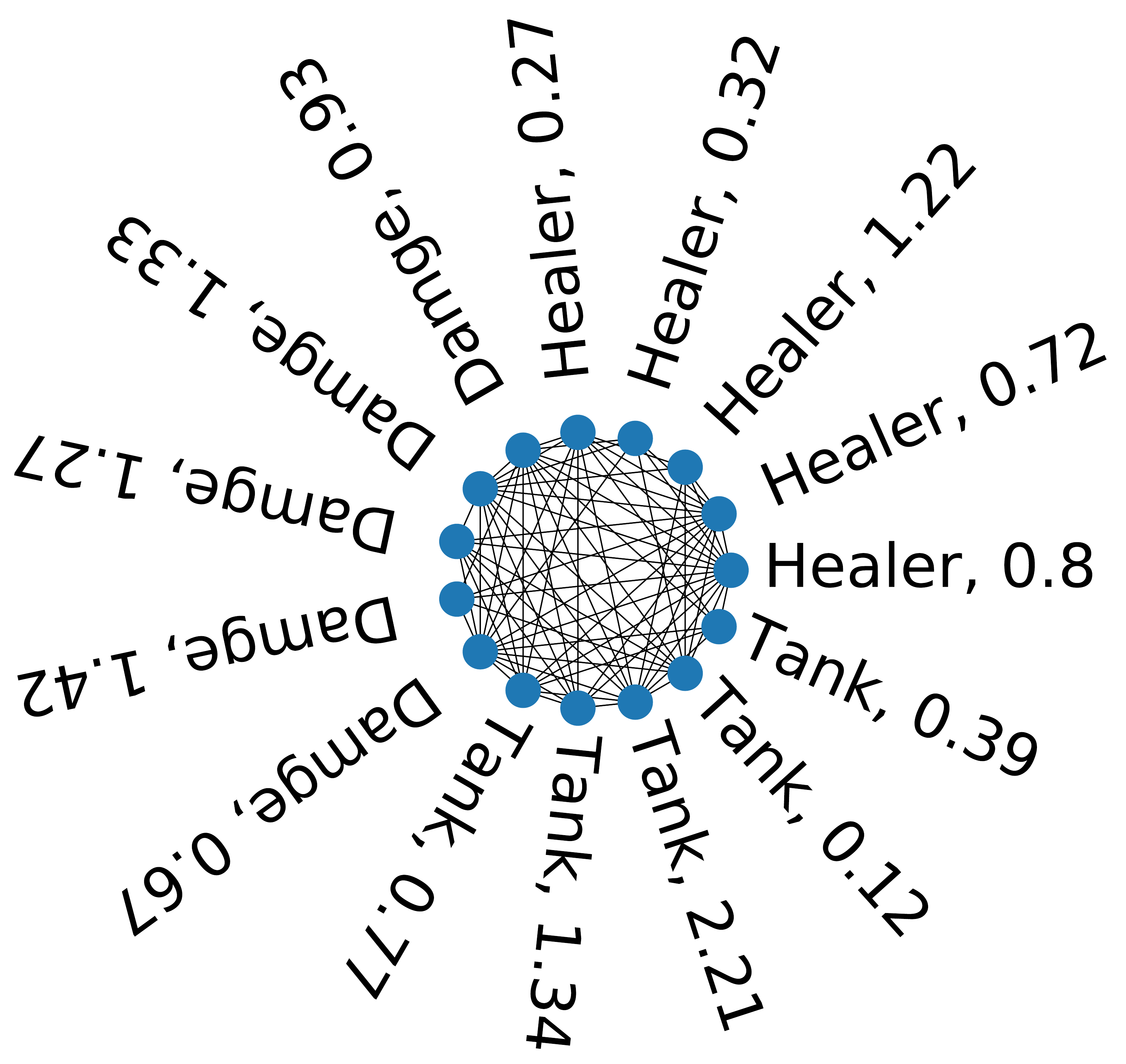}}
     \subfloat[WoW]{\includegraphics[width=3.2cm]{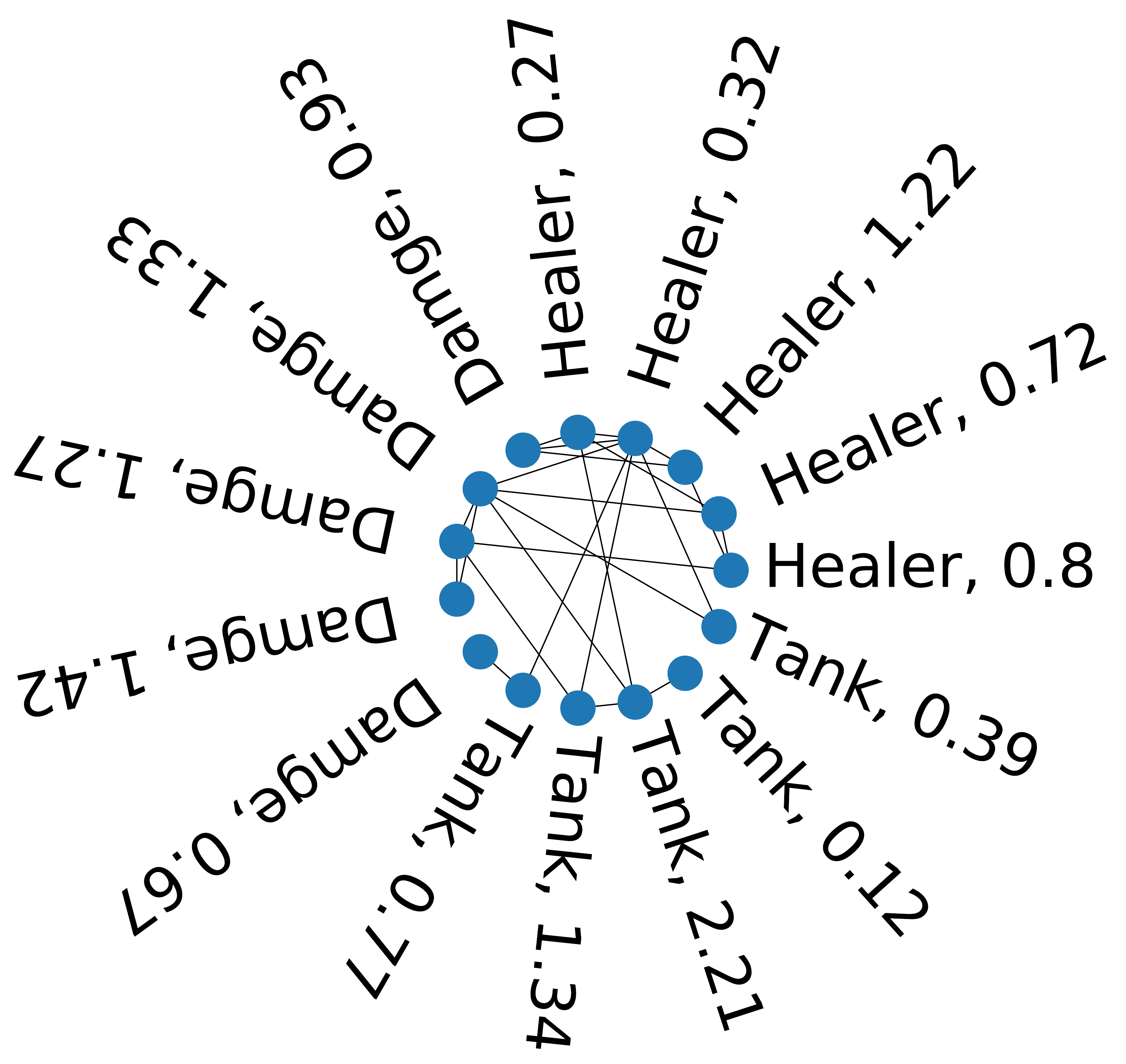}}
     \caption{Formulating large team}\label{fig:lt}
\end{figure}

\noindent\textbf{Small team formulation}. We use the mission location as $\lambda$, $\qk=\{Preiest, Rogue, Hunter, Mage, Warrior \}$, $\rho=1$,  $c=4$, and the players data we collected. The team with 5 players found by \textsf{MKASG} is shown in Figure~\ref{fig:st}(a). First of all, it ensures each suggested class for finishing the mission is in the team. Secondly, the players are near to the location, i.e, 0.21 at most. At last, the relationships between the player are very close.
In comparison, the team formed by the system only ensures the class requirement and players are close to the location. However, the friendships between the players are loose.

\noindent\textbf{Large team formulation}. We use the mission location as $\lambda$, $\qk=\{Damage, Healer, Tank \}$, $\rho=5$,  $c=6$, and the players data we collected. The team with 15 players found by our method and generated by the system are displayed in  Figures~\ref{fig:lt}(a) and (b).
Both of the teams containing team members that are close to the mission and satisfy role requirement for the mission. However, the social relationships of the team found by our method is substantially denser than the social relationships of the team found by the system.

\vspace{-5pt}
\section{Related Works}\label{sec:rw}


\noindent\textbf{Geo-social group discovery}.
Doytsher et al.~\cite{doytsher2010querying}  combined spatial and social networks and proposed graph-based query processing techniques.
Liu et al.~\cite{liu2012circle} proposed a circle-of-friend query to find minimal-diameter social groups.
Yang et al. ~\cite{yang2012socio} considered a special socio-spatial group query with the requirement of minimizing the total spatial distance.
Armenatzoglou et al.~\cite{armenatzoglou2013general} proposed a general framework for geo-social query processing, which separates the social, geographical and query processing modules.
Li et al.~\cite{7079464} studied a geo-social query that retrieves a group of socially connected users whose familiar regions collectively cover a set of query points.
Zhang et al.~\cite{Zhang:2013:IPG:2525314.2525339} proposed a geo-social location recommendation system based on personalized social and geographical influence modeling.
Similarly, Shi et al.~\cite{shi2014density} proposed to cluster and categorize locations based on social and spatial density obtained from geo-social networks.
All these works considered loose social constraints in the query but did not consider keyword cohesiveness.

\noindent\textbf{Team formulation}. Studies on the formation of teams of socially close experts from a social network have drawn additional research interest recently.
However, these studies have mostly focused on minimizing some social metrics in a team without consdering the spatial factor.
Lappas et al.~\cite{lappas2009finding} found a team that covers the required skills and minimizes the structure diameter of the team or the total edge weight of the spanning tree within the team.
In Kargar et al.~\cite{kargar2013finding}, the authors considered forming a team with minimized communication and team costs. However, only the experts who are responsible for at least one required skill are considered in the team cost, and thus cannot be directly applied to our \textsf{MKASG} search problem.
Shen et al.~\cite{shen2016spatial} aimed to find a team that covers appropriate keywords and is spatially close to a location, where the minimum social acquaintance of the team member has not been considered.

\noindent\textbf{Spatial-aware community search}. In~\cite{zhang2017engagement}, they found ($k,r$)-core community such that socially the vertices in $(k,r)$-core is a $k$-core and from similarity perspective pairwise vertices similarity is more than a threshold $r$.
Recently, Three kinds of CS queries have been studied on geo-socialnetworks, namely spatial-aware community search~\cite{FangSpatial}, radius-bounded k-core search~\cite{wang2018efficient}, and geo-social group queries with minimum acquaintance constraint~\cite{zhu2017geo, 7202872}. They all required that the communities are structurally and spatially cohesive.
But, they did not consider textual cohesiveness w.r.t. a set of query keywords as our proposed approach did.

\vspace{-5pt}

\section{Conclusion}\label{sec:con}
In this paper, we study geo-social group search with multi-constraint.
We propose novel search framework making the search towards optimum result fast.
In addition, we propose online data structures, keyword aware union-find structure and keyword-aware forest, which lead the time complexity of  basic search framework to be optimal.
We also propose heuristics, and truss union operation to further speed up the proposed search algorithm.
Extensive experiments are conducted on both synthetic and real datasets, from which the efficiency and the effectiveness are evaluated and justified.




\bibliographystyle{abbrv}
\bibliography{vldb_main}

\end{document}